\documentclass[english]{article}
\usepackage[utf8]{inputenc}
\usepackage[english]{babel}
\usepackage[table,xcdraw]{xcolor}
\usepackage{ulem}
\usepackage{geometry}
\usepackage{float}
\usepackage{amsmath}
\usepackage{graphicx}
\usepackage{natbib}
\usepackage{lipsum}
\usepackage{hyperref}

\floatstyle{ruled}
\newfloat{algorithm}{tbp}{loa}
\providecommand{\algorithmname}{Algorithm}
\floatname{algorithm}{\protect\algorithmname}

\newcommand{\lyxaddress}[1]{
	\par {\raggedright #1
	\vspace{1.4em}
	\noindent\par}
}

\begin{document}

\title{Photodetachment and Test-Particle Simulation Constraints on Negative Ions in Solar System Plasmas}

\author{{Ravindra T. Desai$^1$}, Zeqi Zhang$^1$, Xinni Wu$^1$, Charles Lue$^2$}

\date{}
\maketitle

\vspace{-2em}
\lyxaddress{\begin{center}$^1$Blackett Laboratory, Imperial College London, London, UK \par\end{center}}

\vspace{-2em}
\lyxaddress{\begin{center}$^2$Swedish Institute of Space Physics, Kiruna, Sweden\par\end{center}}

\vspace{-2em}


\begin{abstract}

Negative ions have been detected in abundance in recent years by spacecraft across the solar system. These detections were, however, made by instruments not designed for this purpose and, as such, significant uncertainties remain regarding the prevalence of these unexpected plasma components. In this article, the phenomenon of photodetachment is examined and experimentally and theoretically derived cross-sections are used to calculate photodetachment rates for a range of atomic and molecular negative ions subjected to the solar photon spectrum. These rates are applied to negative ions outflowing from Europa, Enceladus, Titan, Dione and Rhea and their trajectories are traced to constrain source production rates and the extent to which negative ions are able to pervade the surrounding space environments. Predictions are also made for further negative ion populations in the outer solar system with Triton used as an illustrative example. This study demonstrates how, at increased heliocentric distances, negative ions can form stable ambient plasma populations and can be exploited by future missions to the outer solar system.

\end{abstract}

\section{Introduction}

\hspace{2em} Negative ions have long been known to exist in astrophysical environments and within the Earth's ionosphere. They are the major source of opacity within the Sun's outer layers \citep{Wildt39}, appreciate to high densities within interstellar clouds, prestellar cores, and protostellar envelopes \citep[e.g.][]{McCarthy06,Cordiner13} and can outnumber free electrons in the Earth's D region \citep{Appleton33,Pavlov13}.  The Giotto, Galileo, Cassini, Rosetta and Maven spacecraft have found evidence of negative ions existing in abundance within planetary and cometary plasmas with observations at 1P/Halley \citep{Chaizy91}, Europa \citep{Volwerk01}, Titan \citep{Coates07}, Enceladus \citep{Coates10a}, Rhea \citep{Teolis10}, Dione \citep{Nordheim20}, 67/P Churymov/Gerasimenko \citep{Burch15},
Mars \citep{Halekas15} and Saturn's ionosphere \citep{Morooka19}. These observations were, however, serendipitously made with instruments neither designed nor calibrated for this purpose, and the extent to which negative ions pervade space plasmas is not well constrained. This article describes the phenomenon of photodetachment and examines how this process limits the abundance of negative ions that can exist in solar system collisionless plasmas. 

Stable negative ions are produced when an atom or molecule with a positive electron affinity (EA) gains an excess electron. The dominant reactions for producing negative ions in the gas phase include dissociative electron attachment in regions with high fluxes of suprathermal electrons such as at 1P/Halley and Titan \citep{Vuitton09, Cordiner14}, radiative electron attachment reactions in much colder interstellar environments \citep{Herbst81}, and electron- and neutral-induced collisional attachment. Negative ions can also be produced from dust grains and solid surfaces via charge inversion due to scattering, secondary emission and sputtering \citep{Wekhof81,Domingue14}.
The dominant negative ion loss mechanisms  include photodetachment, ion-ion recombination, and ion-neutral associative detachment. Proton transfer, polmerisation and charge exchange reactions can also be efficient in producing and destroying negative ions.  Crucially, the majority of these processes result from collisional reactions with other ions or neutrals with the exceptions of polar photodissociation and photodetachment. Polar photodissociation is a relatively minor reaction channel and within the solar system the reaction rates are many orders of magnitude lower than those of photodetachment. For example, the cross section for the production of H$^-$ from H$_2$ via this process is of the order of 10$^{-23}$ cm$^{2}$ \citep[][]{Chupka75} whereas the photodetachment cross-sections are six orders of magnitude higher, of the order of 10$^{-17}$ cm$^{2}$, see Section \ref{Photodetachment}.  Negative ion populations are therefore unable to be sustained within collisionless environments with high photon fluxes.

This theory is consistent with the early identifications of negative ions.  Within the Sun's atmosphere, H$^-$ is produced within a narrow radial range where the production processes dominate, and indeed do so within all cool stellar photospheres with temperatures less than 7000 K \citep{Wildt39,Chandrasekhar43}. Similarly, negative ions have also been observed by sounding rockets and radar observations of the D and lower E region of the Earth's ionosphere \citep[e.g.][]{Johnson58,Nevejans82}, where collisional reactions enable predominantly halogen- and water-based negative ions to appreciate during the night \citep{Pavlov13}. In the more distant astrophysical environments the reduced photon flux enables radiative electron attachment to efficiently produce a range of carbon-based negative ions \citep{Dalgarno73,Herbst81}.

The recent negative ions detections in the solar system have been made at the interfaces between neutral material and space plasmas. In the inner Coma of 1P/Halley the Giotto spacecraft's electron spectrometer detected negative ion masses consistent with water- and carbon-based molecules \citep{Chaizy91} and at 67P/Churymov-Gerasimenko the Rosetta spacecraft's electron spectrometer detected negative hydrogen ions produced through the interaction between the solar wind and cometary neutrals \citep{Burch15}. Similarly, the MAVEN spacecraft detected negative hydrogen ions from the Martian upper atmosphere \citep{Halekas15}.
 At Jupiter's moon Europa, the Galileo spacecraft's magnetometer observed evidence of a negative ion plasma instability which was used to infer the presence of outflowing negative chlorine ions \citep{Volwerk01,Desai17b}. Within the Saturn system, the Cassini spacecraft's electron spectrometer detected negative carbon- and/or nitrogen-based ions in Titan's ionosphere extending up to 13,800 amu/q \citep{Coates07,Coates09,Wellbrock13,Desai17a,Wellbrock19,Mihailescu20}, which Cassini's Langmuir Probe found to constitute up to 96\% of the ionospheric negative charge density \citep{Wahlund09,Agren12,Shebanits13,Shebanits16}.  Water-based negative ions and dust were similarly detected within the heart of the Enceladus plumes \citep{Coates10a,Coates10,Jones09,Haythornthwaite20} where they also carried the majority of the negative charge density \citep{Morooka11}. Further oxygen and possibly carbon-based negative ions populations have also been identified outflowing from Rhea \citep{Teolis10,Desai18} and Dione \citep{Nordheim20,Tokar12}. Negative ions and dust have also been shown to dominate in Saturn's deep ionosphere and, as at Titan and Enceladus, carry the majority of the negative charge and therefore the majority of the planet's ionospheric plasma mass density due to their large inferred masses \citep{Morooka19,Shebanits20}. 

This article is laid out as follows; the introduction provides a brief review of negative ions in the solar system and where they have been previously detected. Section \ref{Photodetachment} then reviews the phenomenon of photodetachment and photodetachment rates are calculated for a number of atomic and molecular negative ion species. Section \ref{Tracing} then uses test-particle simulations to examine how these rates constrain the negative ion detections already made and their subsequent spatial and temporal evolution. Several predictions are also made for further negative ion species likely to be detected. Section \Ref{Summary} then discusses and summarises the key conclusions and makes the case for dedicated negative ion sensors to be carried on future missions to the outer solar system.

\begin{figure*}[ht]
\includegraphics[width=0.6\textwidth]{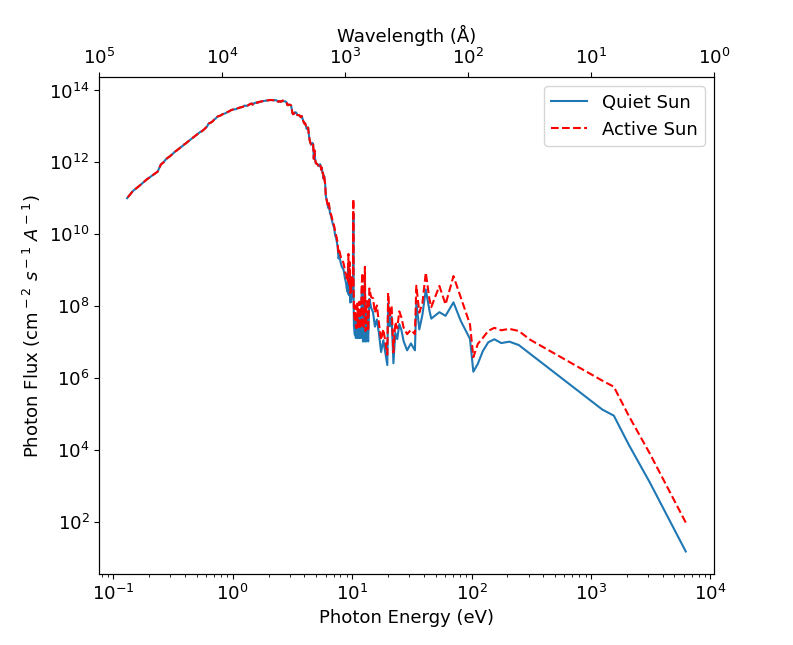}
\centering
\caption{Solar photon flux density for the active and the quiet Sun \citep{Huebner92}. The upper x-axis shows this as a function of photon wavelength and the corresponding photon energy is shown on the lower axis.}
\label{fig1}
\end{figure*}

\section{Photodetachment}
\label{Photodetachment}

\begin{figure*}
\centering
\includegraphics[width=1.0\textwidth]{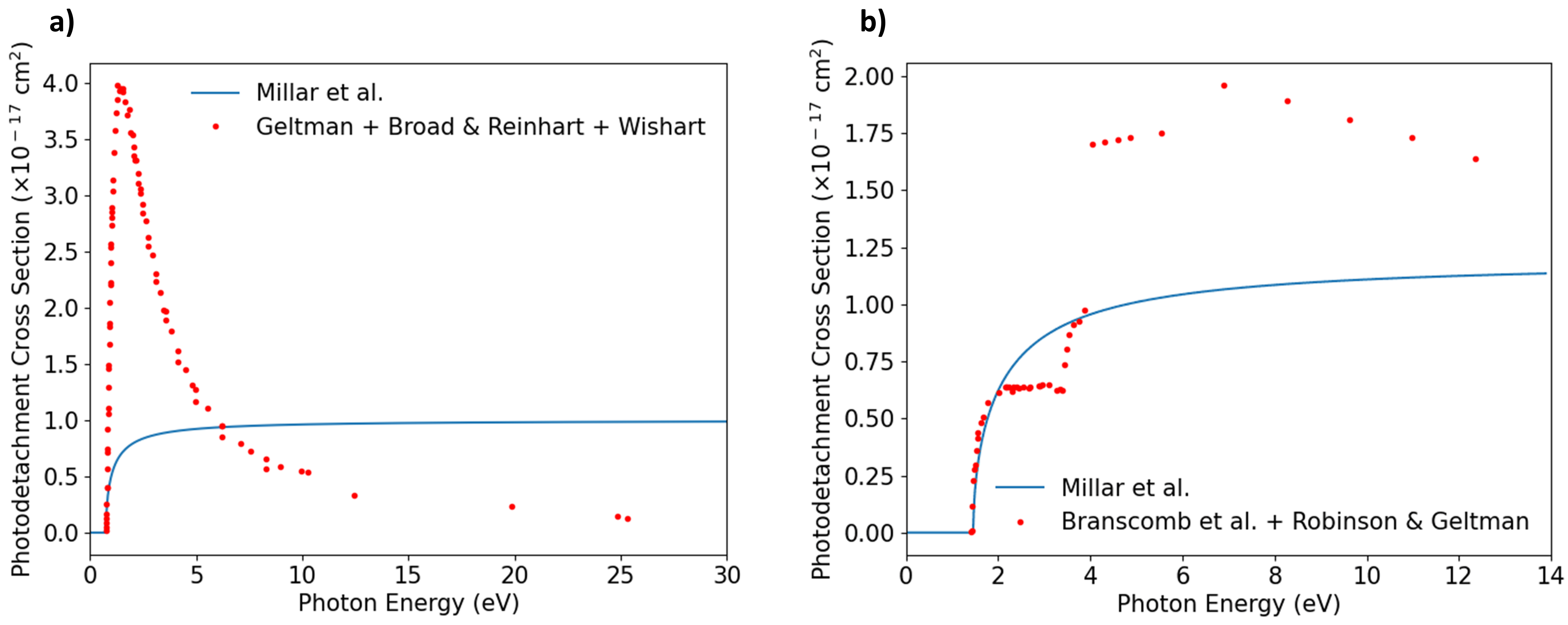}
\caption{Photodetachment cross-sections of (a) H$^-$ and (b) O$^-$ compared to the model of \citet{Millar07} in Equation \ref{emp}.
}
\label{fig2}
\end{figure*}  

\subsection{Theory}
\hspace{2em} Negative ions possess a greater number of electrons than protons, which results in an overall negatively charged atomic ion or molecule. 
The stability of this negative ion is determined by the EA of the neutral atom \citep{Mulliken34}. This is a quantitative measure of the amount of energy required to remove an electron from the negative ion. This determines the energy threshold and reaction rates of such a process, and consequently the half-life of a negative ion.
The EA can either be a positive or negative quantity and only atoms or molecules with a positive EA are
able to exist in a stable negatively charged state. Approximately 80\% of elements in the periodic table have been discovered to possess a positive EA.

Photodetachment is a process where a photon is absorbed by an electron within a negative ion. If the energy of the photon is greater than the electron affinity of the negative ion, the excited electron will have enough energy to be emitted by the atom. In the solar system, the most common source of energetic photons is from the Sun. Figure \ref{fig1} shows the solar flux of the active and the quite Sun from \citet[][and references therein]{Huebner92}, both of which exhibit similar distributions with most of the photons having energies less than ten electron volts, enough to neutralise the majority of negative ions. As a result, electrons that are absorbed by such a photon will have enough energy to be ejected via the reaction sequence,

\begin{equation}
\label{eq1}
A^- + hv \rightarrow  [A^-]^* \rightarrow A + e^-,
\end{equation}
where A denotes a candidate negative ion, $hv$ denotes a photon, and the $*$ denotes an unstable excited state of the negative ion prior to emitting the excess electron.

\subsection{Method}

In this study, we use the method described by \citet{Huebner92} to calculate the photodetachment rates of a range of atomic and molecular negative ions. Ignoring attenuation through mediums, the rate coefficient $k(\lambda_i)$ resulting from the solar flux's wavelength between $\lambda_i$ and $\lambda_i+\Delta \lambda$ is given by
\begin{equation}
    k_i(\lambda_i) = \int_{\lambda_i}^{\lambda +\Delta \lambda_i} \sigma(\lambda)\Phi(\lambda) e^{-\tau(\lambda)} d\lambda,
\end{equation}
where $\sigma(\lambda)$ is the photodetachment cross section of the negative ion and $\Phi(\lambda)$ is the solar flux at wavelength $\lambda$. However, neither $\sigma$ or $\Phi$ is recorded as a continuous function during observations. If the bin width for $\sigma(\lambda)$ is relatively narrow, the above equation is well approximated by
\begin{equation}
    k_i(\lambda_i) =  \sigma_i \Phi_i( \lambda_i ),
\end{equation}
where $\sigma_i$ denotes the average photodetachment cross section between  $\lambda_i$ and $\lambda_i+\Delta \lambda$. $\Phi_i( \lambda_i )$ is represented by the given photon flux between $\lambda_i$ and $\lambda_i+\Delta \lambda $ as
\begin{equation}
    \Phi_i( \tau_i ) = \int_{\lambda_i}^{\lambda +\Delta \lambda_i} \sigma(\lambda) e^{-\tau(\lambda)} d\lambda.
\end{equation}
The total rate coefficient is thus expressed as 
\begin{equation}
    k(\tau) = \sum_{i}^{} k_i (\tau_i).
\end{equation}
The solar spectrum shown in Figure \ref{fig1} peaks between 1--10 eV ( 10$^4$--10$^3$ \r{A}), before decreasing by several orders of magnitude to photon energies near to 10$^4$ eV ($\approx$1 \r{A}). Differences between the quiet and active Sun only become apparent at $>$10 eV (10$^3$ \r{A}) and therefore do not affect the photodetachment calculations within the accuracy which they can be performed. 

The absolute measurement of photodetachment cross-sections of various negative ions is an ongoing research effort \citep[e.g.][]{Lee79,Best11} with drift-tubes, guided ion beams, supersonic beams, and ion traps successfully employed to this end \citep[e.g.][]{Mikosh10}. However, much of the emphasis for these experiments has been to calculate the near-threshold behaviour of the photodetachment cross sections, as well as measuring the asymptotic values for larger photon energies. A wide range of experimental data, from near threshold to photon energies of several electron volts, are therefore unavailable for a number of negative ions.

The model described by \citet{Millar07} provides a prediction for photodetachment cross sections via the relation,
\begin{equation}
    \sigma(\epsilon) = \sigma_\infty \sqrt{1-\frac{EA}{\epsilon}},
    \label{emp}
\end{equation}
where $\sigma_\infty$ is the asymptotic cross section for large photon energies of that ion. This assumes a square-root growth near the threshold, and asymptotic behaviour for photons well above the threshold energy. 
The near threshold behaviour has been verified by many experiments since then \citep[e.g.][]{Best11}, however, the asymptotic behaviours well-above threshold do not always match the experimental values. This behaviour is especially significant for smaller ions which have a smaller range of electron binding energies. The photodetachment rates are therefore mostly controlled by the near threshold behaviours, which generally haven't yet asymptoted to near-constant values.  It should also be noted that this model won't capture strong resonances at photon energies close to threshold which cannot be ruled out in the absence of experimental data \citep{Millar07}. 

\subsection{Reaction Rates}

In this section, photodetachment rates of a range of negative ions at 1 AU are calculated and presented in Table \ref{table1}. The cross-sections derive from a wide variety of experimental techniques and theoretical calculations, many of which were conducted over 50 years ago. Photodetachment rates, using the model outlined in Equation \ref{emp} \citep{Millar07}, are also compared to and used as the primary method when cross-sectional data is unavailable. 
To demonstrate the applicability of Equation \ref{emp}, Figure \ref{fig2} shows a full range of cross-sections for the negative hydrogen and oxygen ions compared to those predicted by this relationship.
 The accurate literature derived values, as subsequently described, peak near 2 eV, near 4$\times$10$^{-17}$ cm$^{2}$, before reducing rapidly to less than 5$\times$10$^{-18}$ cm$^{2}$ at larger energies. The model does not capture this early peak instead asymptoting at lower and then higher relative values. On the other hand, larger ions have a broader range of electron binding energies and the cross-sections calculated using Equation \ref{emp} better approximate those of the oxygen ion with comparable values up to $\approx$ 5 eV where the majority of the solar photon flux lies.%

\subsubsection{Hydrogen}

The H$^-$ photodetachment cross-sections are derived from the studies of \citet{Geltman62}, \citet{Broad76} and \citep{Wishart79} which produce values within 1 \% of each other using J-matrix calculations, quantum perturbation theory, and the close-coupling expansion method with Hylleraas-type correlation
terms, respectively.
When integrated across the solar spectrum, these cross-sections produce a calculated photodetachment rate of 14.3 $\pm$ 0.143 s$^{-1}$, which is near the same as that reported by \citet{Huebner92} using cross-sections of \citet{Geltman62} and \citet{Broad76}.
The predicted rate based on Equation \ref{emp} is, however, significantly lower at 3.16 s$^{-1}$, calculated using an EA of 0.754 eV \citep{Lykke91} and assumed asymptotic cross-section of 10$^{-17}$ cm$^{2}$ \citep{Millar17}.

\subsubsection{Water-Group}

To study water-based negative ion chemistry, we consider the O$^-$, OH$^-$ and [H$_2$O-O$_2]^-$ negative ions. Experimentally derived cross-sections for O$^-$ for energies up to 4 eV are reported by \citet{Branscombe65} which agree well with the recent experimental data of \citet{Hlavanka09}. Above 4 eV, the cross-sections are supplemented with the quantum perturbation theoretical calculations of \citet{Robinson67} although variations at these wavelengths only result in negligible changes to the integrated rate. The photodetachment rate of O$^-$ is calculated to be 1.43 $\pm$ 0.07 s$^{-1}$ and the predicted rate using Equation \ref{emp} is 1.51 $\pm$ 0.08 s$^{-1}$, calculated using an EA of 1.46 eV \citep{Chaibi10} and asymptotic cross-section of 1.2$\times$10$^{-17}$ cm$^{2}$ \citep{Millar07}.

Experimentally measured cross-sections for both OH$^-$ and [H$_2$O-O$_2]^-$ are taken from the study of \citet{Lee79} and the OH$^-$ cross-sections also agree well with the recent experiments of \citet{Hlavanka09}. 
The resultant OH$^-$ photodetachment rate is calculated to be 1.69 $\pm$ 0.25 s$^{-1}$ which is similar to O$^-$. Equation \ref{emp} however predicts a larger photodetachment rate of 2.48 $\pm$ 0.38, calculated using an EA of 1.83 eV \citep{Smith97} and asymptotic cross-section of 3.3$\times$10$^{-17}$ cm$^{2}$ \citep{Hlavanka09}.
The photodetachment rates of the larger [H$_2$O-O$_2]^-$ negative ion are notably lower than these smaller molecules at 0.0688 $\pm$ 0.0172 s$^{-1}$. Due to the lack of information on a precise EA and reduced cross-sectional measurements, we do not calculate a rate for this molecule using Equation \ref{emp}. 

\subsubsection{Halogens}

Halogens possess the largest electron affinities in the periodic table and here we consider Cl$^-$, Br$^-$, and F$^-$. Photodetachment cross-sections for these molecules are provided by the quantum perturbation calculations of \citet{Robinson67} which agree well with the experiments of \citet{Berry63}. As the theoretical calculations do not report errors these are estimated from these previous experiments. The Cl$^-$ rate is calculated to be 0.113$^{+0.0904}_{-0.0527}$ s$^{-1}$, the Br$^{-}$ rate 0.235 $\pm$ 0.141 s$^{-1}$ and the F$^-$ rate 0.0692 $\pm$ 0.0419 s$^{-1}$. 
To calculate the associated rates from Equation \ref{emp}, we take asymptotic cross-sections from \citet{Robinson67} and electron affinities from \citet{Berry63}. The resultant rates fall within the uncertainties of those calculated using the experimentally validated cross-sections. 

\subsubsection{Carbon-based}

To represent carbon-based negative ion chemistry, we consider C$^-$ and the molecular negative ions, CN$_x^-$ (where x=1,3,5) and C$_x$H$^-$ (where x=2,4,6).  The C$^-$ cross-sections are taken from \citet{Seman62} and the  resultant photodetachment rate is 4.06 $\pm$ 0.29 s$^{-1}$. The estimated rate from Equation \ref{emp} is lower at 3.28 $\pm$ 0.29 s$^{-1}$. The asymptotically measured cross-sections are available for five of the molecular negative ions considered \citep{Best11,Kumar13}, but not C$_5$N$^-$ for which $\sigma_\infty$ is assumed to be 10$^{-17}$ cm$^{2}$ \citep{Millar07}. Due to the lack of experimental data across the full range of energies for these molecules, the \citet{Millar07} model is used instead. The calculated rates become smaller with increasing mass with the C$_2$H$^-$ rate calculated at 0.0796 $\pm$ 0.0199 s$^{-1}$, the CN$^-$ rate 0.0399 $\pm$ 0.0010 s$^{-1}$, the C$_4$H$^-$ rate 0.0234 $\pm$ 0.00585 s$^{-1}$, the C$_3$N$^-$ rate 0.0208 $\pm$ 0.0106 s$^{-1}$,  the C$_6$H$^-$ rate 0.00799 $\pm$ 0.00110 s$^{-1}$ and the C$_5$N$^-$ rate 0.00248 s$^{-1}$.

\begin{table}[ht]
 \caption{Photodetachment rates calculated using estimated and experimental cross-sections for various atomic and molecular negative ions at 1 au. These rates are scaled using an inverse square law to calculate the photodetachment rates used in Section \ref{Tracing}. }
\hspace{-3em}
\begin{tabular}{|
>{\columncolor[HTML]{FFFFC7}}c 
>{\columncolor[HTML]{FFFFC7}}c 
>{\columncolor[HTML]{FFFFC7}}c 
>{\columncolor[HTML]{FFFFC7}}c 
>{\columncolor[HTML]{FFFFC7}}c 
>{\columncolor[HTML]{FFFFC7}}c |}
\hline
\textbf{\begin{tabular}[c]{@{}c@{}}Negative Ion\\ Species\end{tabular}} & \textbf{\begin{tabular}[c]{@{}c@{}}Mass\\ {[}amu{]}\end{tabular}} & \textbf{\begin{tabular}[c]{@{}c@{}}Electron\\ Affinity\\ {[}eV{]}\end{tabular}} & \textbf{\begin{tabular}[c]{@{}c@{}}Asymptotic\\Cross-section\\{[}10$^{-17}$cm$^{2}${]}\end{tabular}} & \textbf{\begin{tabular}[c]{@{}c@{}}Estimated \\ Photodetachment Rate \\ {[}s$^{-1}${]}\end{tabular}} & \textbf{\begin{tabular}[c]{@{}c@{}} Accurate\\ Photodetachment Rate \\ {[}s$^{-1}${]}\end{tabular}} \\ \hline
H$^-$                                                                      & 1                                                                 & 0.754 & 1      & 3.16                                                                                                   & 14.3 $\pm$ 0.1                                                                                                  \\
C$^-$                                                                       & 12                                                                & 1.26 & 2    & 3.28 $\pm$ 0.23                                                                       & 4.06 $\pm$ 0.29                                                                                                 \\
O$^-$                                                                       & 16                                                                & 1.46 & 1.2     & 1.51 $\pm$ 0.08                                                                   & 1.43 $\pm$ 0.07                                                                                                 \\
\cellcolor[HTML]{FFFFC7}OH$^-$                                             & 17                                                                & \cellcolor[HTML]{FFFFC7}1.83  & 3.3 & 2.48 $\pm$ 0.38                                                & 1.69 $\pm$ 0.25                                                                                              \\
\cellcolor[HTML]{FFFFC7}F$^-$                                               & 18                                                                & \cellcolor[HTML]{FFFFC7}3.40   & 1 &  0.0430                                               & 0.0692 $\pm$ 0.0419                                                                                          \\
\cellcolor[HTML]{FFFFC7}C$_2$H$^-$                                             & 25                                                                & \cellcolor[HTML]{FFFFC7}3.02   & 0.88                                               & 0.0796  $\pm$ 0.0199 & --                                                                                                \\
\cellcolor[HTML]{FFFFC7}CN$^-$                                              & 26                                                                & \cellcolor[HTML]{FFFFC7}3.86  & 2.84                                                & 0.0399 $\pm$ 0.0010 & --                                                                                            \\
Cl$^-$                                                                      & 35                                                                & 3.61     & 4 & 0.108$^{+0.086}_{-0.050}$                                                                     & 0.113$^{+0.090}_{-0.053}$                                                                                                \\
\cellcolor[HTML]{FFFFC7}C$_4$H$^-$                                             & 49                                                                & \cellcolor[HTML]{FFFFC7}3.56   & 0.77                                               & 0.0234 $\pm$ 0.0059 & --                                                                                              \\
\cellcolor[HTML]{FFFFC7}C$_3$N$^-$                                             & 50                                                                & \cellcolor[HTML]{FFFFC7}4.30 & 5.19                                                 & 0.0208 $\pm$ 0.0106 & --                                                                                               \\
\cellcolor[HTML]{FFFFC7}[O$_2$-H$_2$O]$^-$                                    & 50                                                                & \cellcolor[HTML]{FFFFC7}-- & -- & --                                                     & 0.0688 $\pm$ 0.0172                                                                                              \\
\cellcolor[HTML]{FFFFC7}C$_6$H$^-$                                             & 73                                                                & \cellcolor[HTML]{FFFFC7}3.80   & 0.48                                               & 0.00799 $\pm$ 0.00110 & --                                                                                              \\
\cellcolor[HTML]{FFFFC7}C$_5$N$^-$                                             & 74                                                                & \cellcolor[HTML]{FFFFC7}4.50   & 1                                               & 0.00248   & --                                                                                           \\
Br-                                                                     & 80                                                                & 3.36     & 5                                          & 0.234 $\pm$ 0.141                            & 0.235 $\pm$ 0.141                                                                                                \\
 \hline
\end{tabular}
 \label{table1}
\end{table}
\vspace{-1.3em}
\hspace{-2.5em}
\begin{footnotesize}
{\textbf{Note.} Source code to calculate photodetachment rates as a function of the solar spectrum is available at the Zenodo repository: \url{http://doi.org/10.5281/zenodo.4670382} \citep{ZhangData}} 
\end{footnotesize}

\section{Test-Particle Simulations}
\label{Tracing}
\subsection{Overview}
This section uses test-particle tracing simulations to calculate the trajectories of negative ions outflowing from the various bodies where they have been detected. These are combined with the photodetachment rates calculated in Table \ref{table1} to constrain their temporal and spatial evolution. The calculations therefore provide a prediction for where they will survive to, and for what fluxes might be detected by future missions or within existing datasets. They also provide context and an estimate of source densities for previous detections already accomplished.

\begin{figure*}[ht]
\centering
 \includegraphics[width=0.5\textwidth]{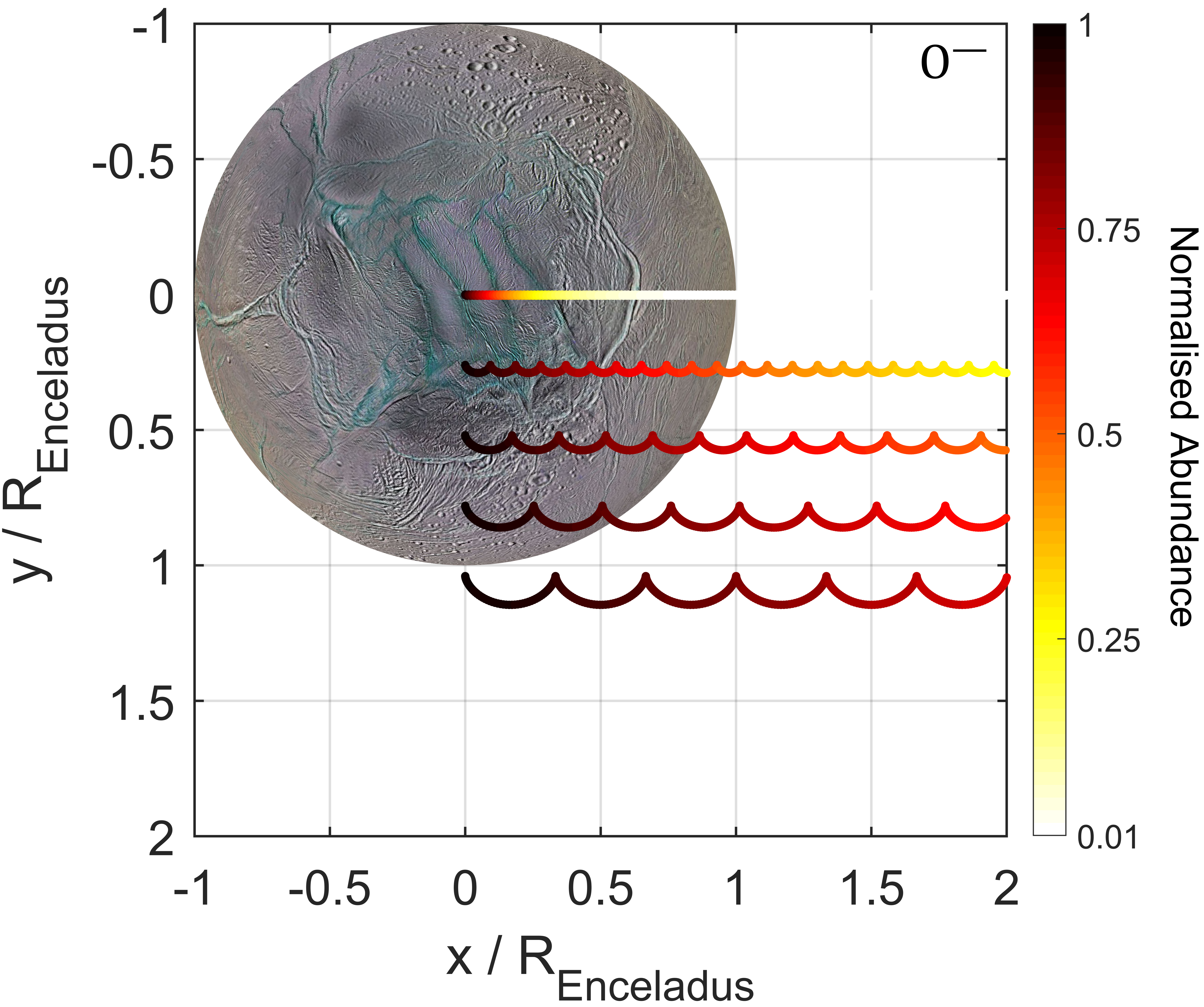}
\caption{O$^-$ outflowing from the Enceladus plumes. The trajectories are coloured according to their normalised maximum source density.  R$_{Enceladus}$ = 252 km..
\label{fig3}}
\end{figure*}

Icy moons constitute the majority of the moons of the outer planets and the detection of sub-surface oceans at Europa, Ganymede and Enceladus, has provided key target environments for future missions such as by the ESA JUpiter ICy moon Explorer (JUICE) mission \citep{Grasset13} and the NASA Europa Clipper mission \citep{Phillips14}. Icy moons typically possess water-based surfaces and transient exospheres, both of which provide information on the bulk and trace compositions of the moons themselves and their sub-surface oceans.
Figures \ref{fig3}--\ref{fig5} show negatively charged oxygen ions outflowing from Europa, Enceladus, Dione and Rhea, ordered by heliocentric distance, and discussed in order of when the observations were made. 

Heavier negative ions have been detected in the inner coma of Comet Halley \citep{Chaizy91}, in Titan's ionosphere \citep{Coates07} and inferred to be outflowing from Europa \citep{Volwerk01}. Table \ref{table1} shows that more massive negative ions possess larger electron affinities and are therefore less susceptible to photodetachment processes. To examine heavier negative ions, Figures \ref{fig6}--\ref{fig8} show chlorine outflowing from Europa and CN$^-$ outflowing from Titan as well as Triton which serves to demonstrate negative ions' increased longevity at increased heliocentric distances.

\subsection{Method}

Newly created negative ions at the interface of space plasma and neutral material will be subjected to the electric field $\mathbf{E = -v \times B}$, where $\mathbf{v}$ is the bulk plasma velocity and $\mathbf{B}$ is the ambient magnetic field. The newly formed negative ion will be accelerated, or `picked up', and gyrate about the magnetic field in accordance with the Lorentz force, and drift in the direction of the bulk plasma flow. The trajectories of newly formed ion populations are calculated in a body-centred cartesian coordinate system where the x-axis points along the direction of plasma flow, the y-axis towards the planet which the body of interest is orbiting, and the z-axis completes the right-handed set. We use the \citet{Boris70} integration technique to evolve particle trajectories assuming nominal dipolar and interplanetary magnetic fields, and locally measured or predicted plasma velocities. The negative ions are initialised at or up to a given altitude above a target body to represent either surface or atmospheric production. For bodies with a substantial ionosphere which can withstand the incoming plasma flow, the plasma flow upstream is approximated as running perpendicularly to this boundary.  Because of these assumptions, ions picked up on an upstream hemisphere are still able to access downstream regions.  The electric field therefore results in negative ions being able to escape from a full given hemisphere of a body immersed within a plasma flow.

The photodetachment rates are applied instantaneously at each calculation timestep. The finite number of particles simulated mean that each particle can be thought of as a macroparticle representing a large collection of particles as in the particle-in-cell approximation \citep{Hockney81}. Due to the variation in the measured abundances at the various bodies during different flybys and the instrument uncertainties involved, the particle fluxes are presented as normalised to the maximum density and specific densities and rates discussed alongside in text. The calculated trajectories are therefore generically applicable to existing datasets as well as future observations that might measure different densities and escape rates. 

\subsection{Enceladus}
\label{Enceladus}
The Cassini spacecraft performed close flybys of Enceladus on multiple occasions and sampled the plumes discovered to be erupting from its south pole \citep{Dougherty06}. The electron spectrometer detected negatively charged water group ions \citep{Coates10a} and dust grains \citep{Jones09} when passing through on flybys E3, E5, E17 and E18 \citep{Hill12,Dong15} and signatures of which were also seen by the Langmuir Probe \citep{Morooka11}. The negative ion masses were identified as distinct peaks in the electron spectrum which were consistent with multiples of 16-18 amu up to 500 amu/q, indicating negative ions derived from water and water-group clusters, such as O$^-$, OH$^-$, and heavier negative water-group ions. The energy resolution of the instrument was optimised for electrons and therefore did not permit more precise mass constraints on these negative ions.

The photodetachment rates of OH$^-$ are similar to that for O$^-$ and, heavier negative water-group ions, approximated here by [O$_2$-H$_2$O$]^-$, have photodetachment rates an order of magnitude smaller than these, thus revealing water cluster ions as especially long-lived. The Cassini detections were made deep inside the plume where the plasma had slowed considerably and the negative ions were suggested to be produced from the low speed thermal gas emission and travelling at low velocities of 0.2--2 km/s  \citep{Haythornthwaite20} . 
The density of the low mass negative ions of mass range 9--27 amu were observed in the range of 1 -- 100 cm$^{-3}$. The densities of negative ions in the constrained range of 45--70 amu, likely comprising of water cluster ions, were detected at lower densities of 0.1 -- 1 cm$^{-3}$ \citep{Coates10a}. These densities were however calculated using an instrument detection efficiency of 5 \% and subsequent studies \citep{Desai18,Nordheim20} use an increased rates of $\approx$20 \%, derived from the studies of \citet{Peko00} and \citet{Stephen00}. The negative ion densities are therefore likely lower than reported by \citet{Coates10a} but with large uncertainties. The uncertainties in the photodetachment rates can therefore be seen to be small compared to uncertainties in the measurements themselves.

The trajectories of these negative ions are calculated assuming a Saturnian dipole field with a field strength at the Enceladus orbit of $\approx$325 nT with the particles initialised below the moon's south pole at various distances from a nominal plume centre with plasma velocities ranging from 1 to 26 km/s.  Figure \ref{fig3} shows a view of the Enceladus south pole with negative oxygen ion trajectories traced and coloured according the their depreciating abundances. At the edge of the plume, assuming a nominal corotational plasma velocity, negative oxygen ions are able to persist for tens of Enceladus radii downstream. If they are picked up in the centre of the plume, however, they are only able to exist within a confined spatial extent of a few tens of kilometres. [O$_2$-H$_2$O]$^-$ would however survive for significantly longer than this as the distance travelled downstream can be scaled by the photodetachment rate. 
The Cassini detections were possible due to the spacecraft ram velocity which boosted the energy in the spacecraft frame such that they could be detected by the electrostatic analyser. The trajectories and lifetimes depicted in Figure 3b therefore confirms that, due to their relatively short lifetimes, these negative ions were produced deep within the Enceladus plume. 

Ion cyclotron waves have been observed throughout the Enceladus Neutral Cloud and E-ring \citep{Meeks16} with amplitudes peaking near Enceladus and inferring large quantities of pickup ions \citep{Cowee09}. The intense radiation fluxes near-saturate the plasma detectors \citep{Taylor17} which makes direct pickup ion identification difficult. Ion cyclotron waves therefore present an alternative method for detecting negative pickup ions, as at Europa \citep{Volwerk01}, as the wave polarisation will be left-handed for positive ions and right-handed for negative ions \citep{Desai17a}.

\begin{figure*}[ht]
\centering
 \includegraphics[width=0.9\textwidth]{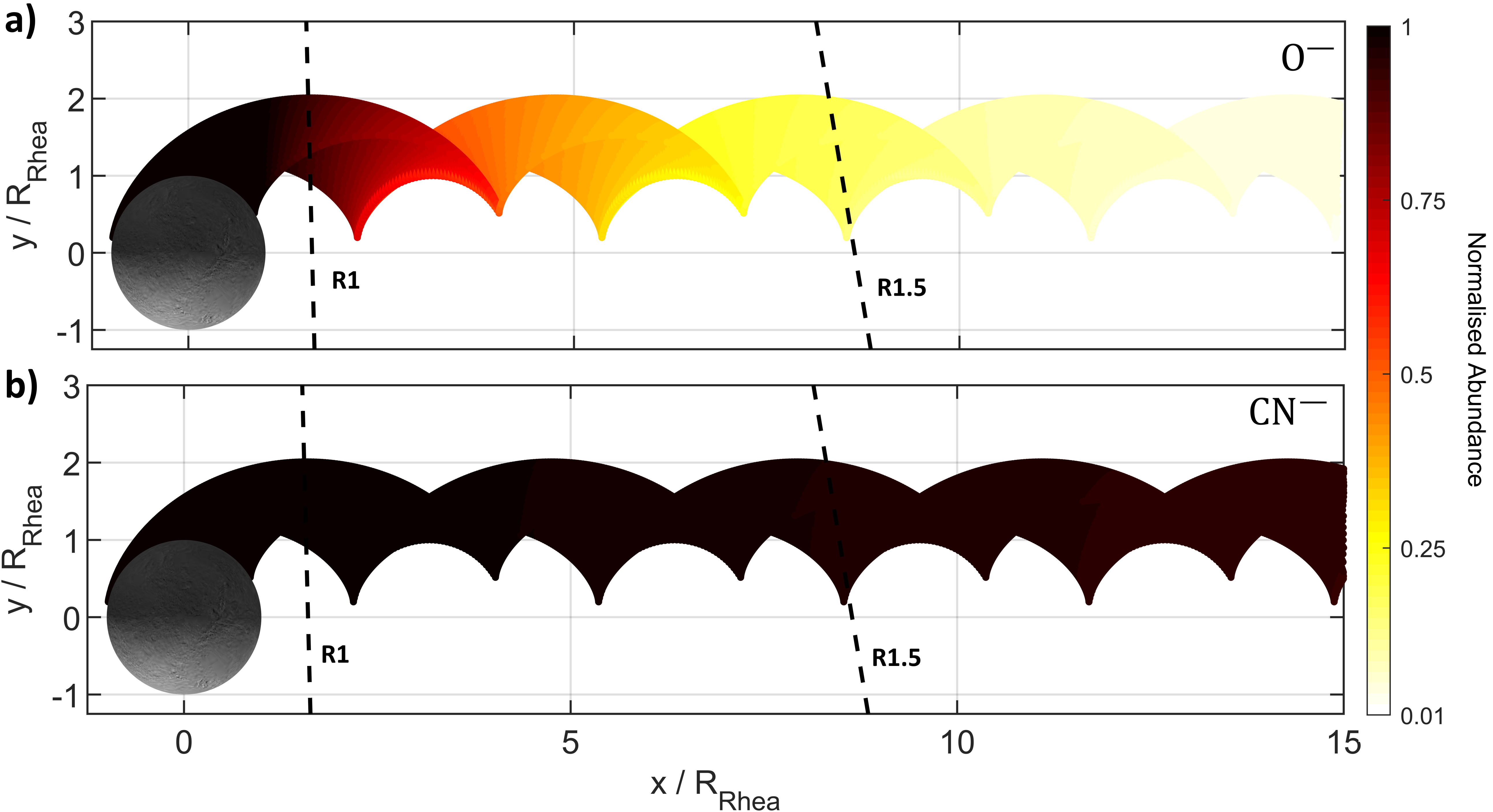}
\caption{O$^-$ and CN$^-$ outflowing from Rhea. The trajectories are coloured according to their normalised maximum source density and the R1 and R1.5 Cassini flybys are marked with dashed blacked lines.  R$_{Rhea}$ = 764 km.
\label{fig4}}
\end{figure*}

\subsection{Rhea}

The detection of negative ions by the Cassini spacecraft at Rhea contributed to the discovery of the moon's sputter-induced oxygen and carbon dioxide exosphere \citep{Teolis10}. 
These negative ions were initially identified as O$^-$ \citep{Teolis10} but their energies, and therefore masses, were later identified as consistent with heavier carbon-based negative ions such as CN$^-$, C$_2$H$^-$, C$_2^-$ or HCO$^-$ \citep{Desai18}. Rhea and other icy moons possess unidentified dark patches at near-infrared wavelengths which has been highlighted as evidence for carbonacious compounds \citep{Scipioni14} from which carbon-based negative ions might originate \citep{Johnson93}. Several processes could however increase oxygen pickup ion energies to this increased energy and, given how little is known about this distant environment, it remains an open question as to which negative ions were detected. The negative fluxes were observed up to densities over 5 $\times$ 10$^{-3}$ cm$^{-3}$ \citep{Desai18}. The uncertainty on this can be reasonably assumed as 17 \% corresponding to the energy resolution of the instrument, which is therefore of a similar magnitude to the error in the O$^-$ photodetachment rate, but smaller than the error in the CN$^-$ photodetachment rate, see Table \ref{table1}.

The trajectories of outflowing O$^-$ and CN$^-$ are calculated assuming a Saturnian dipolar magnetic field of $\approx$26 nT at Rhea's orbit and a plasma velocity of 60 km/s \citep{Wilson10}.  Figure \ref{fig4} and \ref{fig4} shows the resultant trajectories and depreciating abundances for negative ions initialised at Rhea's surface, with illumination and plasma conditions representative of the R1 encounter where they were detected.
These show that the oxygen ions would depreciate to 1 \% of their source densities within 10 Rhea radii and by nearly a quarter by the time they had reached Cassini, 
An oxygen composition therefore corresponds to source production rates $\approx$25 \% higher than those reported of over 6.25 $\times$ 10$^{-3}$ cm$^{-3}$. Carbon-based negative ions would however survive significantly longer, with CN$^-$ surviving up to over 500 Rhea radii downstream before depreciating by the same levels. Carbon-based negative ions would therefore be considered to have suffered negligible losses en-route to Cassini from the moon.

\begin{figure*}[ht]
\centering
 \includegraphics[width=1.0\textwidth]{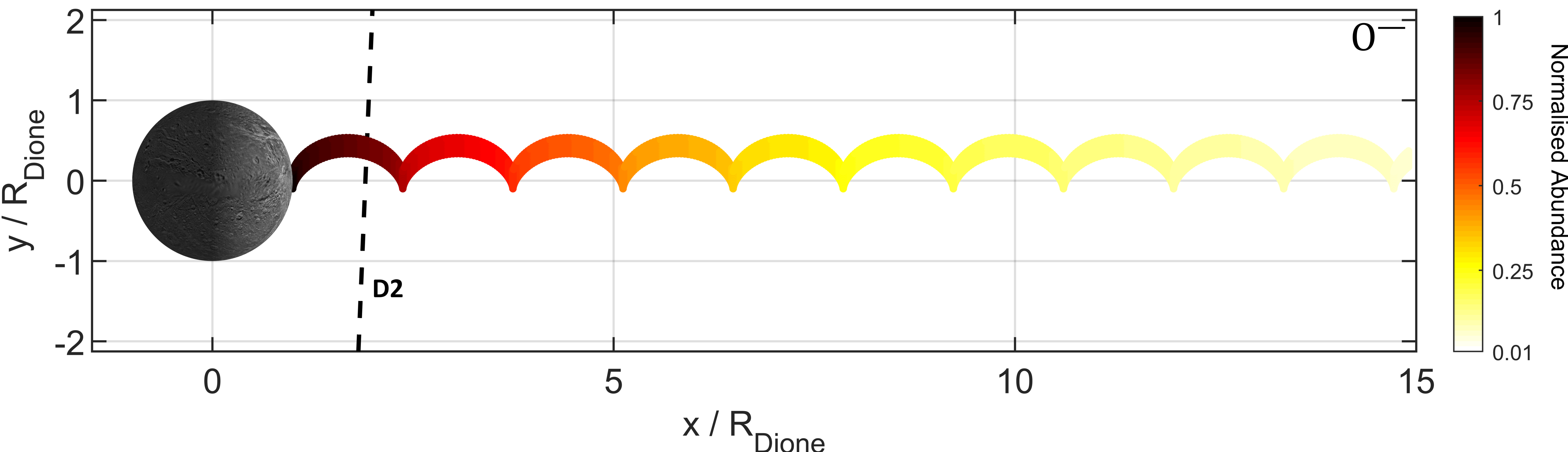}
\caption{O$^-$ and CN$^-$ outflowing from Dione. The trajectories are coloured according to their normalised maximum source density and the D2 Cassini flyby is marked with a dashed blacked line.  R$_{Dione}$ = 561 km.
\label{fig5}}
\end{figure*}

\subsection{Dione}

Dione was discovered to possess a sputtered-induced oxygen and carbon dioxide exosphere \citep{Tokar12}. During D2, the second flyby of Dione, the Cassini spacecraft detected fluxes of outflowing negative ions within the moon's wake consistent with negative ions of masses 15--25 amu/q \citep{Nordheim20}. These were therefore identified as O$^-$  with densities of $\approx 3 \times 10 ^ {-3}$ cm$^{-3}$, comparable to those of the positive ion outflows \citep{Tokar12}.

These negative ion detections were made just 1000 km downstream of the moon and therefore within the particles first gyration around the magnetic field. The particle trajectories shown in Figure \ref{fig5} assume a Saturnian dipole field of $\approx$75 nT at Dione's orbit with particles spawned on Dione's surface where the detections have been identified as originating \citep{Nordheim20}.  The particle trajectories show that negative oxygen ion fluxes will survive to nearly 15 Dione radii downstream before decreasing to 1 \% of their source values. The normalised abundances also show that by the time of the detections during the second Cassini-Dione conjunction D2, photodetachment losses would have been a noticeable loss channel and that the sources fluxes would be approximately 25\% higher than those reaching the Cassini spacecraft, nearer $4 \times 10 ^ {-3}$ cm$^{-3}$. As at Rhea, the uncertainties in these densities can be considered to derive from the 17 \% energy resolution of the instrument, which is of a similar magnitude to the errors in the photodetachment rate. 

\subsection{Europa}
\label{europa1}
\subsubsection{Water-Group}
\begin{figure*}[ht]
\centering
 \includegraphics[width=0.9\textwidth]{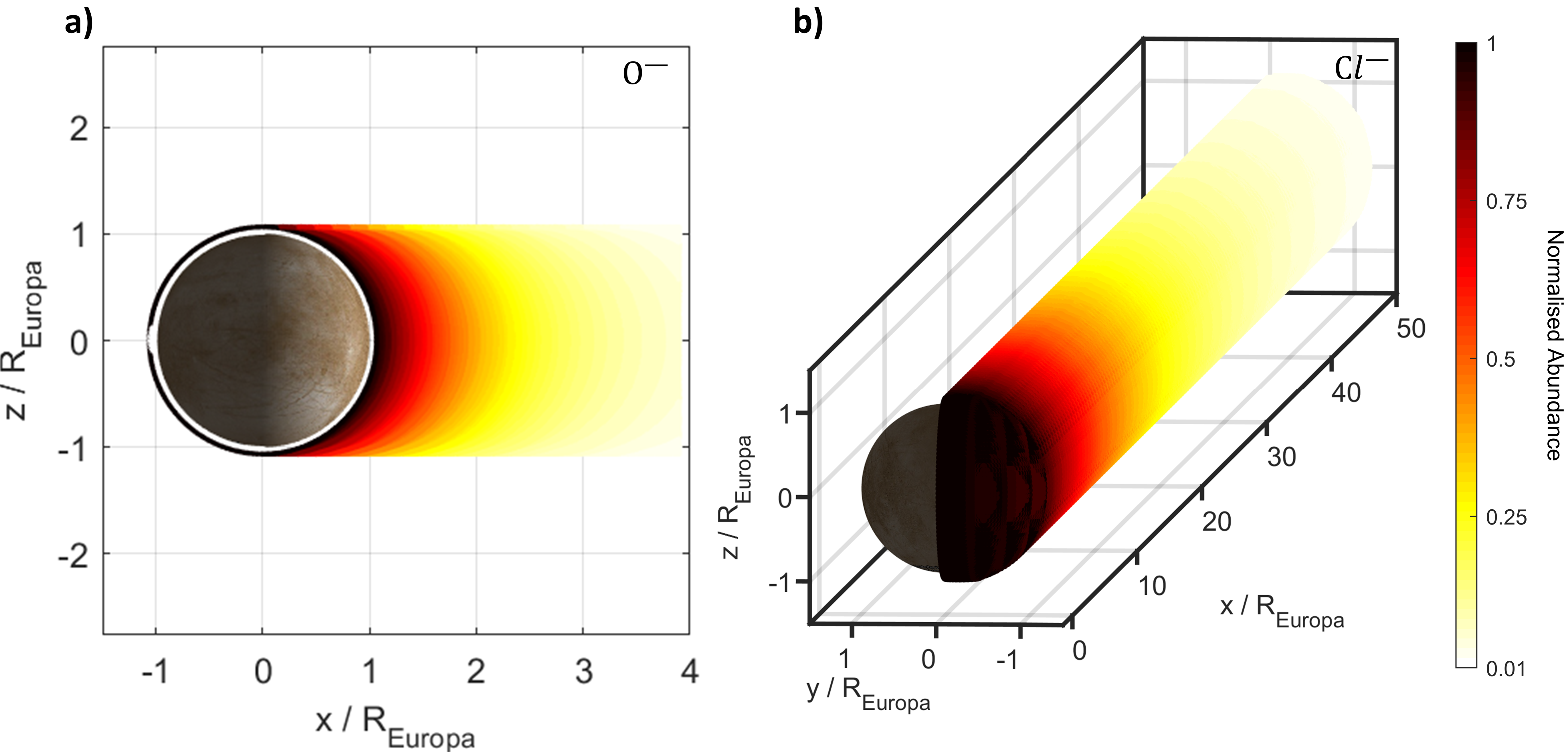}
\caption{ Cl$^-$ outflowing from Europa.The trajectories are coloured according to the normalised maximum density. R$_{Europa}$ = 1561 km.
\label{fig6}}
\end{figure*}

Europa possesses an oxygen exosphere \citep{Hall95} and large sub-surface water ocean \citep{Khurana98}. Hubble observations have also provided tentative plume detections \citep{Roth14,Sparks16} and the Galileo spacecraft observations have been shown to be consistent with their presence \citep{Jia18,Huybrighs}. 
Negative oxygen ions are therefore likely to be produced in a similar fashion to at Enceladus within the putative plumes via electron attachment reactions, or sputtered from the water-ice surface as at Dione, and possibly Rhea, via surface mediated processes resulting from the intense Jovian radiation environment \citep{VanAllen75}. 
The trajectories of outflowing negative oxygen ions are calculated in Figure \ref{fig6} assuming a dipolar magnetic field strength at Europa of $\approx$400 nT with particles initialised only in the visible x-z plane, 100 km above the Europan surface. The trajectories reveal that negative oxygen ions can survive to approximately 4 Europa radii downstream before depreciating by two orders of magnitude. This distance might be lower if the plasma velocity is slowed significantly near the moon, as was sometimes observed \citep{Kivelson09}. Figure \ref{fig3}a shows that the presence of negative oxygen ions, or further water group negative ions which would survive for even longer, could be important in interpreting bulk plasma current measurements from Langmuir Probe or Faraday Cup instruments in this region, depending on their source densities. 

\subsubsection{Chlorine}

Europa's surface is predominantly water-ice but contains dark streaks within cracks in the icy shell which \citet{Hand15} find are consistent with heavily irradiated chlorine compounds originating from the moon's ocean.
Outflowing positive and negative chlorine ions were inferred to exist by the polarisation of ion cyclotron waves \citep{Volwerk01} observed by the Galileo spacecraft in the moon's plasma wake. Hybrid simulations by \citet{Desai17b} support this hypothesis and use the wave amplitudes to calculate source densities in the range of 0.1 -- 1 cm$^{-3}$. 

The trajectories of Cl$^-$ are calculated as outlined in Section \ref{europa1}. Figure \ref{fig6} shows these and reveals that fluxes are able to survive up to $\approx$50 Europan radii downstream before decreasing to 1 \% of their source densities. This is significantly further downstream then where Galileo observed right-handed wave power at the chlorine ion gyrofrequency at just 3-4 Europan radii \citep{Volwerk01}. The photodetachment rates therefore indicate that local fluxes in this region would be representative of the source production, whether globally or locally produced. 
It is also interesting to note the C$_2$H $_x^-$ negative ions are also predicated as a potential sputter product at Europa \citep{Johnson93}.

\begin{figure*}[ht]
\centering
 \includegraphics[width=0.5\textwidth]{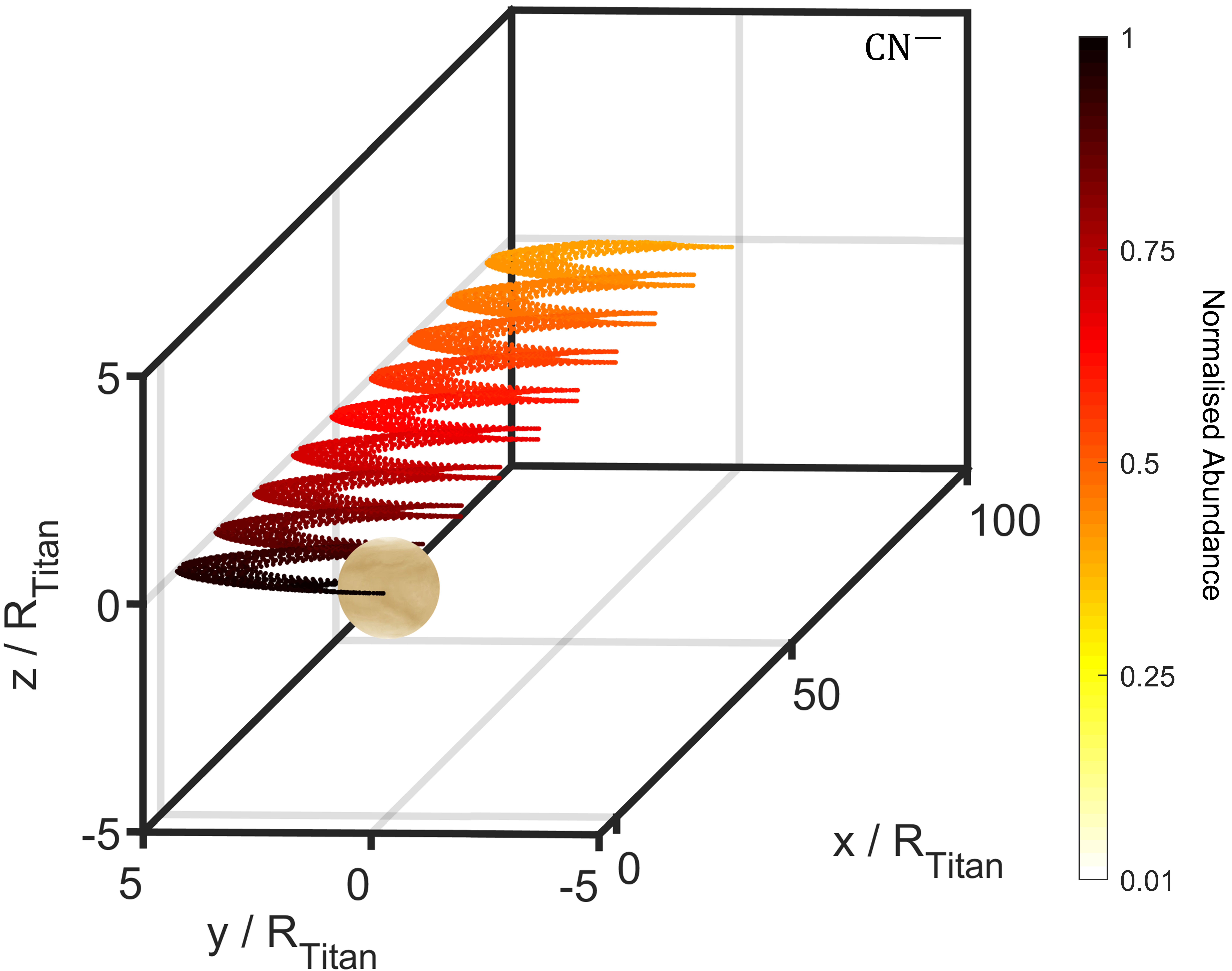}
\caption{(a) CN$^-$ outflowing from Titan. The trajectories are coloured according to the normalised maximum density. R$_{Titan}$ = 2576 km. 
\label{fig7}}
\end{figure*}

\subsection{Titan}

Titan's ionosphere contains an abundance of carbon- and nitrogen-based ion and neutral chemistry, including, negative ions with masses up to 13,800 amu/q \citep{Coates07,Coates09}. Within the broad mass spectrum distinct peaks were identified at 25.8--26.0 and 49.0-50.1 amu/q belonging to the carbon chain anions CN$^-$/C$_3$N$^-$ and/or C$_2$H$^-$/C$_4$H$^-$ \citep{Desai18}, which extend right up to near Titan's exobase near 1400 km. Chemical models predict these carbon chain anions \citep{Vuitton09} and also the lighter negative hydrogen ion H$^-$ \citep{Mukundan18} although this is below the lower end of the mass detection threshold of Cassini's spectrometer and so was not detected. 

Global hybrid simulations of Titan's plasma interaction \citep{Ledvina12} were found to be highly dependent on the negative ions within Titan's ionosphere. These simulations predict large quantities of  20-70 amu negative ions escaping from the ionosphere and being picked up and drifting downstream. In Figure \ref{fig7}, the trajectories of these negative ions are calculated assuming Saturnian dipolar magnetic fields of $\approx$6 nT at Titan's orbit and a bulk plasma velocity of 100 km/s, with particles initialised only at the equatorial regions of Titan to visually highlight the large gyroradius relative to the moon. 
Figure \ref{fig7} shows negative ions fluxes are able to survive for extended distances and reach $\approx$300 Titan radii downstream before  depreciating to 1 \% of their source densities.

Larger carbon-based negative ions are also predicted to escape and will survive for even greater distances due to their lower photodetachment rates, see Table \ref{table1}. While such negative pickup ions haven't been detected as yet, positive pickup ions have been \citep{Regoli16} and Figure \ref{fig7} therefore provides a prediction for where outflowing negative ion fluxes would persist to.

\subsection{Triton}

Triton possesses a nitrogen and methane hazy atmosphere \citep{Broadfoot89} with highly electronegative molecules like at Titan. Photochemical processes have been observed as instigating Titan's negative ion chemistry through the production of suprathermal electron populations which produce Titan's low-mass negative ions \citep{Mihailescu20}. Despite the reduced solar photon flux at Triton's $\approx$30 au heliocentric distance, the Voyager 2 spacecraft observed a large ionosphere which was explained by the influx of magnetospheric plasma \citep{Hoogeveen94} with sufficient energies to produce negative ions such as CN$^-$ \citep{Vuitton09,Mihailescu20}.

\begin{figure*}[ht]
\centering
 \includegraphics[width=0.6\textwidth]{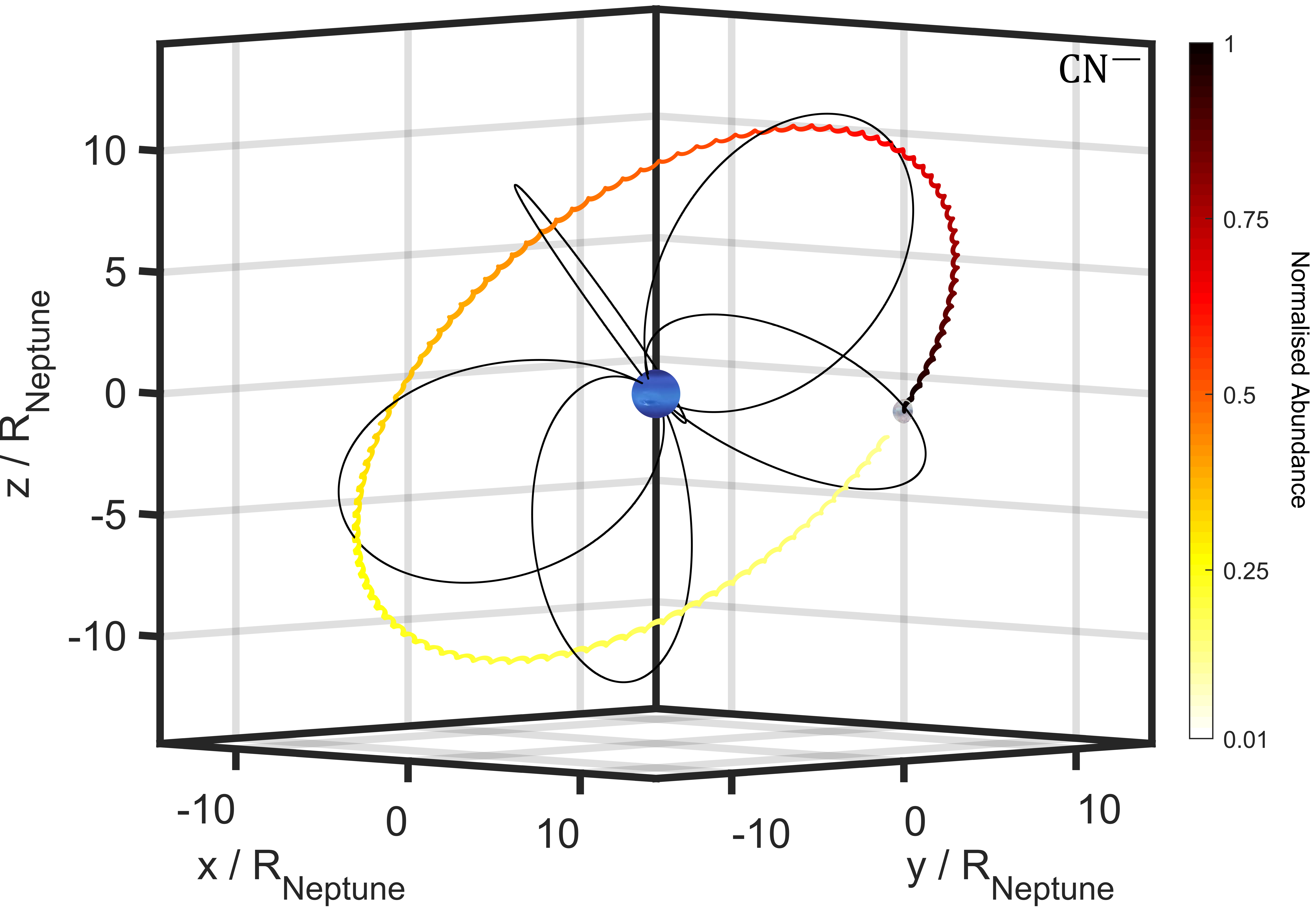}
\caption{CN$^-$ outflowing from Triton. The trajectories are coloured according to the normalised maximum density. R$_{Neptune}$ = 24622 km. Neptune is to scale but Triton is magnified to make visible. 
\label{fig8}}
\end{figure*}

The Neptunian dipole is highly inclined and rotating and the dynamics of trapped particles populations are surely a research end in itself. The trajectories of outflowing CN$^-$ are calculated within a magnetospheric configuration of a dipole field inclined 45$^{\circ}$ relative to Triton's orbital plane, with a Neptunian dipole moment of 2.2$\times$10$^{17}$ Tm$^{-3}$.  Figure \ref{fig8} shows the traced abundances of the candidate CN$^-$ negative ion and shows that negative ions such as CN$^-$ would form a full torus of negative ions over 24 hours in magnetic local time before the fluxes depreciate to 1 \% of their source densities.

The low photodetachment rates at these heliocentric distances reveal negative ions as capable of forming stable magnetospheric populations which can be used to probe the composition of the Neptunian and Uranian moons, as well as transport processes within these complex magnetospheric configurations. 

\section{Discussion \& Conclusions\label{sec:Conclusion-and-Future}}
\label{Summary}

This study was motivated by the spate of negative ion detections in solar system plasmas over the past thirty years. The method of \citet{Huebner92} was used to calculate the photodetachment rates of a range of atomic and molecular negative ions subjected to the solar photon spectrum, as presented in Table \ref{table1}.  Significant effort was made in deriving as complete and accurate set of cross-sectional data as possible, much of which was derived from experimental studies conducted over fifty years ago. The results of these calculations were compared to photodetachment rates calculated using the theoretical relationship of \citet{Millar07}. This approach proved accurate for larger molecules, as long as an accurate electron affinity and asymptotic cross-section at larger photon energies had been determined. For the smaller molecules examined, H$^-$, C$^-$, O$^-$, and OH$^-$, this method, however, proved to be inaccurate.
The newly calculated photodetachment rates spanned four orders of magnitude with decreasing photodetachment rates corresponding to increasing mass, a trend which can be attributed to the increased electron affinity of the larger molecules negating regions of the solar spectrum where the photon flux is highest. 

The photodetachment rates were then scaled by heliocentric distance and test-particle simulations used to constrain the nominal escape trajectories of negative ion populations at various bodies in the outer solar system. Negative water-group ions are constrained at Europa and Enceladus, negative oxygen ions at Rhea and Dione, negative chlorine ions at Europa, and negative carbon-based ions at Titan and Rhea. These detections were made with instruments not specifically designed for detecting negative ions and significant uncertainties therefore exist on the reported densities. For this reason, the escaping fluxes were normalised to the source densities thus leaving the results as broadly applicable to existing as well as future measurements.

The photodetachment rates of negative ions derived from water-based negative ion chemistry provided several interesting insights. O$^-$, OH$^-$, and negative water cluster ions such as [O$_2$-H$_2$O]$^-$, were identified as presenting a viable detection aim at the Jovian icy moons, with heavier negative cluster ions being particularly resistant to photodetachment processes. These rates also had interesting implications in the Saturn system. At Dione, the detected negative ions were shown to have already undergone significant photodetachment losses and the source negative oxygen ion production rates are therefore inferred to be $\approx$25 \% greater than those measured in-situ \citep{Nordheim20}. This would also be true for O$^-$ populations detected at Rhea \citep{Teolis10} although ambiguities remain as to whether the detected negative ions consisted of O$^-$ or heavier carbon-based species \citep{Desai18} which would have undergone negligible photdetachment losses. It was also found that the stagnated plasma flow in the Enceladus plume mean that locally detected water group ions \citep{Coates10a,Haythornthwaite20} are confined to a limited spatial domain.
 
Heavier negative ions possess increased electron affinities and the corresponding photodetachment rates provided several further interesting insights. At Europa, negative chlorine ion densities were found to persist for up to 50 Europa radii downstream and sufficiently long to power the wave activity at the chlorine gyrofrequency observed by Galileo \citep{Volwerk01,Desai17b}, but not long enough to contribute to the Europan plasma torus \citep{Bagenal15}. At Titan, CN$^-$ escaping from the moon's extended ionosphere could survive for $\approx300$ Titan radii downstream and are thus predicted to form an extended negative ion tail.They are therefore also able to constitute a significant part of the moon's plasma interaction as predicted by \citet{Ledvina12}. 
 
In addition to these aforementioned locations, negatively charged hydrogen has been detected at 67P/Churymov-Gerasimenko and at Mars, produced via double charge exchange with the solar wind \citep{Burch15,Halekas15}, and likely do at many, if not all, unmagnetised, bodies interacting directly with the solar wind. The persistence of such fluxes will be strongly controlled by photodetachment which will evolve as such during their highly elliptical orbits. 

The negative ion populations considered have all been found to exist within collisionless solar system plasmas surrounding moons, comets and Mars. The equatorial ionosphere of Saturn has, however, also been found to contain large populations of negatively charged ions and dust which comprise up to over 95 \% of the negative charge density \citep{Morooka19}. These have significant consequences for the ionospheric dynamics \citep{Shebanits20} and the spacecraft-plasma interaction experienced by Cassini \citep{Zhang21}.
Saturn's polar wind has been observed as a significant mass source for its magnetosphere \citep{Felici16} and if negatively charged ions and aerosols exist within the polar regions, these will also mass load the magnetosphere with negative ion populations. While these will be relatively short lived outside of Saturn's shadow, polar winds at Uranus and Neptune might produce long-lived ambient magnetospheric populations of negative ions. 

Of further mention is that negative ions will persist within collisionless plasmas if not exposed to direct sunlight. The trajectories presented in Figures \ref{fig3} and \ref{fig4} are calculated for predominantly sunlit conditions and it should therefore be noted that for specific body-planet geometries negative ions might persist for significantly longer. At 9 am planetary magnetic local time an orbiting body's plasma wake will be completely in shadow and negative ions escaping downtail will therefore survive until reaching sunlight. Similarly, if negative ions are produced within the shadow of a planet they will survive for extended periods where losses can only result from infrequent collisions with ions and neutrals, extra-solar photon fluxes or transport into other regions.

While most negative ions have only recently been detected within the solar system, it is worth mentioning that their presence was predicted at bodies with tenuous atmospheres over 40 years ago. \citet{Wekhof81} considered negative ions could be formed at Mercury, the Moon and Europa, via sputtering and charge inversion processes, and constitute a few percent of the negative charge density. Further bodies subjected to similar sputtering processes, such as asteroids, further moons and Mercury \citep{Domingue14}, therefore also present environments where negative ions might persist.

The photodetachment rates are considerably reduced at increased heliocentric distances due to the reduced solar photon fluxes. To illustrate this concept, CN$^-$ fluxes were modelled escaping from Triton's large ionosphere, and were found to survive for 24 hours of magnetic local time thus forming a negative ion torus around Neptune. Future missions to the outer solar system are therefore well-placed to use such populations as a diagnostic tool using dedicated instrumentation \citep[e.g.][]{Lepri17}, whose presence may be necessary to wholly and accurately describe the surrounding environments.

\section*{Acknowledgements}
RTD and CL wish to acknowledge discussions within International Space Science Institute International Team 437: "Negative Ions in the Solar System". ZZ and XW acknowledge the Imperial College First Year Undergraduate Research Projects. Research Projects. RTD receives funding from NERC grant NE/P017347/1.   
\newline

\bibliography{main.bib}

\begin{thebibliography}{94}
\providecommand{\natexlab}[1]{#1}
\providecommand{\url}[1]{\texttt{#1}}
\expandafter\ifx\csname urlstyle\endcsname\relax
  \providecommand{\doi}[1]{doi: #1}\else
  \providecommand{\doi}{doi: \begingroup \urlstyle{rm}\Url}\fi

\bibitem[{{\AA}gren} et~al.(2012){{\AA}gren}, {Edberg}, and {Wahlund}]{Agren12}
K.~{{\AA}gren}, N.~J.~T. {Edberg}, and J.-E. {Wahlund}.
\newblock {Detection of negative ions in the deep ionosphere of Titan during
  the Cassini T70 flyby}.
\newblock \emph{Geophysical Research Letters}, 39:\penalty0 L10201, May 2012.
\newblock \doi{10.1029/2012GL051714}.

\bibitem[{Appleton} et~al.(1933){Appleton}, {Naismith}, and
  {Builder}]{Appleton33}
E.~V. {Appleton}, R.~{Naismith}, and G.~{Builder}.
\newblock {Ionospheric Investigations in High Latitudes}.
\newblock \emph{Nature}, 132:\penalty0 340--341, Sept. 1933.
\newblock \doi{10.1038/132340a0}.

\bibitem[{Arijs} et~al.(1982){Arijs}, {Nevejans}, {Frederick}, and
  {Ingels}]{Nevejans82}
E.~{Arijs}, D.~{Nevejans}, P.~{Frederick}, and J.~{Ingels}.
\newblock {Stratospheric negative ion composition measurements, ion abundances
  and related trace gas detection}.
\newblock \emph{Journal of Atmospheric and Terrestrial Physics}, 44:\penalty0
  681--694, Aug. 1982.
\newblock \doi{10.1016/0021-9169(82)90130-1}.

\bibitem[{Bagenal} et~al.(2015){Bagenal}, {Sidrow}, {Wilson}, {Cassidy},
  {Dols}, {Crary}, {Steffl}, {Delamere}, {Kurth}, and {Paterson}]{Bagenal15}
F.~{Bagenal}, E.~{Sidrow}, R.~J. {Wilson}, T.~A. {Cassidy}, V.~{Dols}, F.~J.
  {Crary}, A.~J. {Steffl}, P.~A. {Delamere}, W.~S. {Kurth}, and W.~R.
  {Paterson}.
\newblock {Plasma conditions at Europa's orbit}.
\newblock \emph{Icarus}, 261:\penalty0 1--13, Nov. 2015.
\newblock \doi{10.1016/j.icarus.2015.07.036}.

\bibitem[{Berry} and {Reimann}(1963)]{Berry63}
R.~S. {Berry} and C.~W. {Reimann}.
\newblock {Absorption Spectrum of Gaseous F- and Electron Affinities of the
  Halogen Atoms}.
\newblock \emph{Journ. of Comp. Phys.}, 38\penalty0 (7):\penalty0 1540--1543,
  Apr. 1963.
\newblock \doi{10.1063/1.1776916}.

\bibitem[{Best} et~al.(2011){Best}, {Otto}, {Trippel}, {Hlavenka}, {von
  Zastrow}, {Eisenbach}, {J{\'e}zouin}, {Wester}, {Vigren}, {Hamberg}, and
  {Geppert}]{Best11}
T.~{Best}, R.~{Otto}, S.~{Trippel}, P.~{Hlavenka}, A.~{von Zastrow},
  S.~{Eisenbach}, S.~{J{\'e}zouin}, R.~{Wester}, E.~{Vigren}, M.~{Hamberg}, and
  W.~D. {Geppert}.
\newblock {Absolute Photodetachment Cross-section Measurements for Hydrocarbon
  Chain Anions}.
\newblock \emph{Astrophysical Journal}, 742\penalty0 (2):\penalty0 63, Dec.
  2011.
\newblock \doi{10.1088/0004-637X/742/2/63}.

\bibitem[{Boris}(1970)]{Boris70}
J.~{Boris}.
\newblock {}.
\newblock In \emph{Proceedings of the Fourth Conference on Numerical Simulation
  of Plasmas}, page~3, June 1970.

\bibitem[Branscomb et~al.(1965)Branscomb, Smith, and Tisone]{Branscombe65}
L.~M. Branscomb, S.~J. Smith, and G.~Tisone.
\newblock Oxygen metastable atom production through photodetachment.
\newblock \emph{The Journal of Chemical Physics}, 43\penalty0 (8):\penalty0
  2906--2907, 1965.
\newblock \doi{10.1063/1.1697230}.
\newblock URL \url{https://doi.org/10.1063/1.1697230}.

\bibitem[{Breyer} et~al.(1981){Breyer}, {Frey}, and {Hotop}]{Smith97}
F.~{Breyer}, P.~{Frey}, and H.~{Hotop}.
\newblock {High resolution photoelectron spectrometry of negative ions:
  Rotational transitions in laser-photodetachment of OH$^{-}$, SH$^{-}$,
  SD$^{-}$}.
\newblock \emph{Zeitschrift fur Physik A Hadrons and Nuclei}, 300\penalty0
  (1):\penalty0 7--24, Mar. 1981.
\newblock \doi{10.1007/BF01412609}.

\bibitem[{Broad} and {Reinhardt}(1976)]{Broad76}
J.~T. {Broad} and W.~P. {Reinhardt}.
\newblock {One- and two-electron photoejection from H$^{ - }$: A multichannel
  J-matrix calculation}.
\newblock \emph{Physical Review A: General Physics}, 14\penalty0 (6):\penalty0
  2159--2173, Dec. 1976.
\newblock \doi{10.1103/PhysRevA.14.2159}.

\bibitem[{Broadfoot} et~al.(1989){Broadfoot}, {Atreya}, {Bertaux}, {Blamont},
  {Dessler}, {Donahue}, {Forrester}, {Hall}, {Herbert}, {Holberg}, {Hunten},
  {Krasnopolsky}, {Linick}, {Lunine}, {Mcconnell}, {Moos}, {Sandel},
  {Schneider}, {Shemansky}, {Smith}, {Strobel}, and {Yelle}]{Broadfoot89}
A.~L. {Broadfoot}, S.~K. {Atreya}, J.~L. {Bertaux}, J.~E. {Blamont}, A.~J.
  {Dessler}, T.~M. {Donahue}, W.~T. {Forrester}, D.~T. {Hall}, F.~{Herbert},
  J.~B. {Holberg}, D.~M. {Hunten}, V.~A. {Krasnopolsky}, S.~{Linick}, J.~I.
  {Lunine}, J.~C. {Mcconnell}, H.~W. {Moos}, B.~R. {Sandel}, N.~M. {Schneider},
  D.~E. {Shemansky}, G.~R. {Smith}, D.~F. {Strobel}, and R.~V. {Yelle}.
\newblock {Ultraviolet spectrometer observations of Neptune and Triton}.
\newblock \emph{Science}, 246:\penalty0 1459--1466, Dec. 1989.
\newblock \doi{10.1126/science.246.4936.1459}.

\bibitem[{Burch} et~al.(2015){Burch}, {Cravens}, {Llera}, {Goldstein},
  {Mokashi}, {Tzou}, and {Broiles}]{Burch15}
J.~L. {Burch}, T.~E. {Cravens}, K.~{Llera}, R.~{Goldstein}, P.~{Mokashi}, C.-Y.
  {Tzou}, and T.~{Broiles}.
\newblock {Charge exchange in cometary coma: Discovery of H$^{-}$ ions in the
  solar wind close to comet 67P/Churyumov-Gerasimenko}.
\newblock \emph{Geophysical Research Letters}, 42:\penalty0 5125--5131, July
  2015.
\newblock \doi{10.1002/2015GL064504}.

\bibitem[{Chaibi} et~al.(2010){Chaibi}, {Pel{\'a}ez}, {Blondel}, {Drag}, and
  {Delsart}]{Chaibi10}
W.~{Chaibi}, R.~J. {Pel{\'a}ez}, C.~{Blondel}, C.~{Drag}, and C.~{Delsart}.
\newblock {Effect of a magnetic field in photodetachment microscopy}.
\newblock \emph{European Physical Journal D}, 58\penalty0 (1):\penalty0 29--37,
  May 2010.
\newblock \doi{10.1140/epjd/e2010-00086-7}.

\bibitem[{Chaizy} et~al.(1991){Chaizy}, {Reme}, {Sauvaud}, {D'Uston}, {Lin},
  {Larson}, {Mitchell}, {Anderson}, {Carlson}, {Korth}, and {Mendis}]{Chaizy91}
P.~{Chaizy}, H.~{Reme}, J.~A. {Sauvaud}, C.~{D'Uston}, R.~P. {Lin}, D.~E.
  {Larson}, D.~L. {Mitchell}, K.~A. {Anderson}, C.~W. {Carlson}, A.~{Korth},
  and D.~A. {Mendis}.
\newblock {Negative ions in the coma of Comet Halley}.
\newblock \emph{Nature}, 349:\penalty0 393--396, Jan. 1991.
\newblock \doi{10.1038/349393a0}.

\bibitem[{Chandrasekhar} and {Krogdahl}(1943)]{Chandrasekhar43}
S.~{Chandrasekhar} and M.~K. {Krogdahl}.
\newblock {On the Negative Hydrogen Ion and its Absorption Coefficient.}
\newblock \emph{Astrophysical Journal}, 98:\penalty0 205, Sept. 1943.
\newblock \doi{10.1086/144561}.

\bibitem[Chupka et~al.(1975)Chupka, Dehmer, and Jivery]{Chupka75}
W.~A. Chupka, P.~M. Dehmer, and W.~T. Jivery.
\newblock High resolution photoionization study of ion‐pair formation in h2,
  hd, and d2.
\newblock \emph{The Journal of Chemical Physics}, 63\penalty0 (9):\penalty0
  3929--3944, 1975.
\newblock \doi{10.1063/1.431833}.
\newblock URL \url{https://aip.scitation.org/doi/abs/10.1063/1.431833}.

\bibitem[{Coates} et~al.(2007){Coates}, {Crary}, {Lewis}, {Young}, {Waite}, and
  {Sittler}]{Coates07}
A.~J. {Coates}, F.~J. {Crary}, G.~R. {Lewis}, D.~T. {Young}, J.~H. {Waite}, and
  E.~C. {Sittler}.
\newblock {Discovery of heavy negative ions in Titan's ionosphere}.
\newblock \emph{Geophysical Research Letters}, 34:\penalty0 L22103, Nov. 2007.
\newblock \doi{10.1029/2007GL030978}.

\bibitem[{Coates} et~al.(2009){Coates}, {Wellbrock}, {Lewis}, {Jones}, {Young},
  {Crary}, and {Waite}]{Coates09}
A.~J. {Coates}, A.~{Wellbrock}, G.~R. {Lewis}, G.~H. {Jones}, D.~T. {Young},
  F.~J. {Crary}, and J.~H. {Waite}.
\newblock {Heavy negative ions in Titan's ionosphere: Altitude and latitude
  dependence}.
\newblock \emph{Planetary and Space Science}, 57:\penalty0 1866--1871, Dec.
  2009.
\newblock \doi{10.1016/j.pss.2009.05.009}.

\bibitem[{Coates} et~al.(2010{\natexlab{a}}){Coates}, {Jones}, {Lewis},
  {Wellbrock}, {Young}, {Crary}, {Johnson}, {Cassidy}, and {Hill}]{Coates10a}
A.~J. {Coates}, G.~H. {Jones}, G.~R. {Lewis}, A.~{Wellbrock}, D.~T. {Young},
  F.~J. {Crary}, R.~E. {Johnson}, T.~A. {Cassidy}, and T.~W. {Hill}.
\newblock {Negative ions in the Enceladus plume}.
\newblock \emph{Icarus}, 206:\penalty0 618--622, Apr. 2010{\natexlab{a}}.
\newblock \doi{10.1016/j.icarus.2009.07.013}.

\bibitem[{Coates} et~al.(2010{\natexlab{b}}){Coates}, {Wellbrock}, {Lewis},
  {Jones}, {Young}, {Crary}, {Waite}, {Johnson}, {Hill}, and
  {Sittler}]{Coates10}
A.~J. {Coates}, A.~{Wellbrock}, G.~R. {Lewis}, G.~H. {Jones}, D.~T. {Young},
  F.~J. {Crary}, J.~H. {Waite}, R.~E. {Johnson}, T.~W. {Hill}, and E.~C.
  {Sittler}, Jr.
\newblock {Negative ions at Titan and Enceladus: recent results}.
\newblock \emph{Faraday Discussions}, 147:\penalty0 293, 2010{\natexlab{b}}.
\newblock \doi{10.1039/c004700g}.

\bibitem[{Cordiner} and {Charnley}(2014)]{Cordiner14}
M.~A. {Cordiner} and S.~B. {Charnley}.
\newblock {Negative ion chemistry in the coma of comet 1P/Halley}.
\newblock \emph{Meteoritics and Planetary Science}, 49:\penalty0 21--27, Jan.
  2014.
\newblock \doi{10.1111/maps.12082}.

\bibitem[{Cordiner} et~al.(2013){Cordiner}, {Buckle}, {Wirstr{\"o}m},
  {Olofsson}, and {Charnley}]{Cordiner13}
M.~A. {Cordiner}, J.~V. {Buckle}, E.~S. {Wirstr{\"o}m}, A.~O.~H. {Olofsson},
  and S.~B. {Charnley}.
\newblock {On the Ubiquity of Molecular Anions in the Dense Interstellar
  Medium}.
\newblock \emph{Astrophysical Journal}, 770:\penalty0 48, June 2013.
\newblock \doi{10.1088/0004-637X/770/1/48}.

\bibitem[{Cowee} et~al.(2009){Cowee}, {Omidi}, {Russell}, {Blanco-Cano}, and
  {Tokar}]{Cowee09}
M.~M. {Cowee}, N.~{Omidi}, C.~T. {Russell}, X.~{Blanco-Cano}, and R.~L.
  {Tokar}.
\newblock {Determining ion production rates near Saturn's extended neutral
  cloud from ion cyclotron wave amplitudes}.
\newblock \emph{Journal of Geophysical Research (Space Physics)}, 114:\penalty0
  A04219, Apr. 2009.
\newblock \doi{10.1029/2008JA013664}.

\bibitem[{Dalgarno} and {McCray}(1973)]{Dalgarno73}
A.~{Dalgarno} and R.~A. {McCray}.
\newblock {The Formation of Interstellar Molecules from Negative Ions}.
\newblock \emph{Astrophysical Journal}, 181:\penalty0 95--100, Apr. 1973.
\newblock \doi{10.1086/152032}.

\bibitem[Desai et~al.(2017{\natexlab{a}})Desai, Coates, Wellbrock, Vuitton,
  Crary, González-Caniulef, Shebanits, Jones, Lewis, Waite, Cordiner, Taylor,
  Kataria, Wahlund, Edberg, and Sittler]{Desai17a}
R.~T. Desai, A.~J. Coates, A.~Wellbrock, V.~Vuitton, F.~J. Crary,
  D.~González-Caniulef, O.~Shebanits, G.~H. Jones, G.~R. Lewis, J.~H. Waite,
  M.~Cordiner, S.~A. Taylor, D.~O. Kataria, J.-E. Wahlund, N.~J.~T. Edberg, and
  E.~C. Sittler.
\newblock Carbon chain anions and the growth of complex organic molecules in
  titan’s ionosphere.
\newblock \emph{The Astrophysical Journal Letters}, 844\penalty0 (2):\penalty0
  L18, 2017{\natexlab{a}}.
\newblock URL \url{http://stacks.iop.org/2041-8205/844/i=2/a=L18}.

\bibitem[Desai et~al.(2017{\natexlab{b}})Desai, Cowee, Wei, Fu, Gary, Volwerk,
  and Coates]{Desai17b}
R.~T. Desai, M.~M. Cowee, H.~Wei, X.~Fu, S.~P. Gary, M.~Volwerk, and A.~J.
  Coates.
\newblock Hybrid simulations of positively and negatively charged pickup ions
  and cyclotron wave generation at europa.
\newblock \emph{Journal of Geophysical Research: Space Physics}, 122\penalty0
  (10):\penalty0 10,408--10,420, 2017{\natexlab{b}}.
\newblock ISSN 2169-9402.
\newblock \doi{10.1002/2017JA024479}.
\newblock URL \url{http://dx.doi.org/10.1002/2017JA024479}.
\newblock 2017JA024479.

\bibitem[Desai et~al.(2018)Desai, Taylor, Regoli, Coates, Nordheim, Cordiner,
  Teolis, Thomsen, Johnson, Jones, Cowee, and Waite]{Desai18}
R.~T. Desai, S.~A. Taylor, L.~H. Regoli, A.~J. Coates, T.~A. Nordheim, M.~A.
  Cordiner, B.~D. Teolis, M.~F. Thomsen, R.~E. Johnson, G.~H. Jones, M.~M.
  Cowee, and J.~H. Waite.
\newblock Cassini caps identification of pickup ion compositions at rhea.
\newblock \emph{Geophysical Research Letters}, pages n/a--n/a, 2018.
\newblock ISSN 1944-8007.
\newblock \doi{10.1002/2017GL076588}.
\newblock URL \url{http://dx.doi.org/10.1002/2017GL076588}.
\newblock 2017GL076588.

\bibitem[{Domingue} et~al.(2014){Domingue}, {Chapman}, {Killen}, {Zurbuchen},
  {Gilbert}, {Sarantos}, {Benna}, {Slavin}, {Schriver},
  {Tr{\'a}vn{\'\i}{\v{c}}ek}, {Orlando}, {Sprague}, {Blewett}, {Gillis-Davis},
  {Feldman}, {Lawrence}, {Ho}, {Ebel}, {Nittler}, {Vilas}, {Pieters},
  {Solomon}, {Johnson}, {Winslow}, {Helbert}, {Peplowski}, {Weider}, {Mouawad},
  {Izenberg}, and {McClintock}]{Domingue14}
D.~L. {Domingue}, C.~R. {Chapman}, R.~M. {Killen}, T.~H. {Zurbuchen}, J.~A.
  {Gilbert}, M.~{Sarantos}, M.~{Benna}, J.~A. {Slavin}, D.~{Schriver}, P.~M.
  {Tr{\'a}vn{\'\i}{\v{c}}ek}, T.~M. {Orlando}, A.~L. {Sprague}, D.~T.
  {Blewett}, J.~J. {Gillis-Davis}, W.~C. {Feldman}, D.~J. {Lawrence}, G.~C.
  {Ho}, D.~S. {Ebel}, L.~R. {Nittler}, F.~{Vilas}, C.~M. {Pieters}, S.~C.
  {Solomon}, C.~L. {Johnson}, R.~M. {Winslow}, J.~{Helbert}, P.~N. {Peplowski},
  S.~Z. {Weider}, N.~{Mouawad}, N.~R. {Izenberg}, and W.~E. {McClintock}.
\newblock {Mercury's Weather-Beaten Surface: Understanding Mercury in the
  Context of Lunar and Asteroidal Space Weathering Studies}.
\newblock \emph{SSR}, 181:\penalty0 121--214, May 2014.
\newblock \doi{10.1007/s11214-014-0039-5}.

\bibitem[Dong et~al.(2015)Dong, Hill, and Ye]{Dong15}
Y.~Dong, T.~W. Hill, and S.-Y. Ye.
\newblock Characteristics of ice grains in the enceladus plume from cassini
  observations.
\newblock \emph{Journal of Geophysical Research: Space Physics}, 120\penalty0
  (2):\penalty0 915--937, 2015.
\newblock ISSN 2169-9402.
\newblock \doi{10.1002/2014JA020288}.
\newblock URL \url{http://dx.doi.org/10.1002/2014JA020288}.
\newblock 2014JA020288.

\bibitem[{Dougherty} et~al.(2006){Dougherty}, {Khurana}, {Neubauer}, {Russell},
  {Saur}, {Leisner}, and {Burton}]{Dougherty06}
M.~K. {Dougherty}, K.~K. {Khurana}, F.~M. {Neubauer}, C.~T. {Russell},
  J.~{Saur}, J.~S. {Leisner}, and M.~E. {Burton}.
\newblock {Identification of a Dynamic Atmosphere at Enceladus with the Cassini
  Magnetometer}.
\newblock \emph{Science}, 311:\penalty0 1406--1409, Mar. 2006.
\newblock \doi{10.1126/science.1120985}.

\bibitem[{Felici} et~al.(2016){Felici}, {Arridge}, {Coates}, {Badman},
  {Dougherty}, {Jackman}, {Kurth}, {Melin}, {Mitchell}, {Reisenfeld}, and
  {Sergis}]{Felici16}
M.~{Felici}, C.~S. {Arridge}, A.~J. {Coates}, S.~V. {Badman}, M.~K.
  {Dougherty}, C.~M. {Jackman}, W.~S. {Kurth}, H.~{Melin}, D.~G. {Mitchell},
  D.~B. {Reisenfeld}, and N.~{Sergis}.
\newblock {Cassini observations of ionospheric plasma in Saturn's magnetotail
  lobes}.
\newblock \emph{Journal of Geophysical Research (Space Physics)}, 121\penalty0
  (1):\penalty0 338--357, Jan. 2016.
\newblock \doi{10.1002/2015JA021648}.

\bibitem[{Geltman}(1962)]{Geltman62}
S.~{Geltman}.
\newblock {The Bound-Free Absorption Coefficient of the Hydrogen Negative Ion.}
\newblock \emph{Astrophysical Journal}, 136:\penalty0 935, Nov. 1962.
\newblock \doi{10.1086/147447}.

\bibitem[{Grasset} et~al.(2013){Grasset}, {Dougherty}, {Coustenis}, {Bunce},
  {Erd}, {Titov}, {Blanc}, {Coates}, {Drossart}, {Fletcher}, {Hussmann},
  {Jaumann}, {Krupp}, {Lebreton}, {Prieto-Ballesteros}, {Tortora}, {Tosi}, and
  {Van Hoolst}]{Grasset13}
O.~{Grasset}, M.~K. {Dougherty}, A.~{Coustenis}, E.~J. {Bunce}, C.~{Erd},
  D.~{Titov}, M.~{Blanc}, A.~{Coates}, P.~{Drossart}, L.~N. {Fletcher},
  H.~{Hussmann}, R.~{Jaumann}, N.~{Krupp}, J.~P. {Lebreton},
  O.~{Prieto-Ballesteros}, P.~{Tortora}, F.~{Tosi}, and T.~{Van Hoolst}.
\newblock {JUpiter ICy moons Explorer (JUICE): An ESA mission to orbit Ganymede
  and to characterise the Jupiter system}.
\newblock \emph{Planetary and Space Science}, 78:\penalty0 1--21, Apr. 2013.
\newblock \doi{10.1016/j.pss.2012.12.002}.

\bibitem[{Halekas} et~al.(2015){Halekas}, {Lillis}, {Mitchell}, {Cravens},
  {Mazelle}, {Connerney}, {Espley}, {Mahaffy}, {Benna}, {Jakosky}, {Luhmann},
  {McFadden}, {Larson}, {Harada}, and {Ruhunusiri}]{Halekas15}
J.~S. {Halekas}, R.~J. {Lillis}, D.~L. {Mitchell}, T.~E. {Cravens},
  C.~{Mazelle}, J.~E.~P. {Connerney}, J.~R. {Espley}, P.~R. {Mahaffy},
  M.~{Benna}, B.~M. {Jakosky}, J.~G. {Luhmann}, J.~P. {McFadden}, D.~E.
  {Larson}, Y.~{Harada}, and S.~{Ruhunusiri}.
\newblock {MAVEN observations of solar wind hydrogen deposition in the
  atmosphere of Mars}.
\newblock \emph{Geophysical Research Letters}, 42\penalty0 (21):\penalty0
  8901--8909, Nov. 2015.
\newblock \doi{10.1002/2015GL064693}.

\bibitem[{Hall} et~al.(1995){Hall}, {Strobel}, {Feldman}, {McGrath}, and
  {Weaver}]{Hall95}
D.~T. {Hall}, D.~F. {Strobel}, P.~D. {Feldman}, M.~A. {McGrath}, and H.~A.
  {Weaver}.
\newblock {Detection of an oxygen atmosphere on Jupiter's moon Europa}.
\newblock \emph{Nature}, 373:\penalty0 677--679, Feb. 1995.
\newblock \doi{10.1038/373677a0}.

\bibitem[{Hand} and {Carlson}(2015)]{Hand15}
K.~P. {Hand} and R.~W. {Carlson}.
\newblock {Europa's surface color suggests an ocean rich with sodium chloride}.
\newblock \emph{Geophysical Research Letters}, 42:\penalty0 3174--3178, May
  2015.
\newblock \doi{10.1002/2015GL063559}.

\bibitem[{Haythornthwaite} et~al.(2020){Haythornthwaite}, {Coates}, {Jones},
  and {Waite}]{Haythornthwaite20}
R.~P. {Haythornthwaite}, A.~J. {Coates}, G.~H. {Jones}, and J.~H. {Waite}.
\newblock {Fast and Slow Water Ion Populations in the Enceladus Plume}.
\newblock \emph{Journal of Geophysical Research (Space Physics)}, 125\penalty0
  (2):\penalty0 e27591, Feb. 2020.
\newblock \doi{10.1029/2019JA027591}.

\bibitem[{Herbst}(1981)]{Herbst81}
E.~{Herbst}.
\newblock {Can negative molecular ions be detected in dense interstellar
  clouds}.
\newblock \emph{Nature}, 289:\penalty0 656, Feb. 1981.
\newblock \doi{10.1038/289656a0}.

\bibitem[{Hill} et~al.(2012){Hill}, {Thomsen}, {Tokar}, {Coates}, {Lewis},
  {Young}, {Crary}, {Baragiola}, {Johnson}, {Dong}, {Wilson}, {Jones},
  {Wahlund}, {Mitchell}, and {Hor{\'a}nyi}]{Hill12}
T.~W. {Hill}, M.~F. {Thomsen}, R.~L. {Tokar}, A.~J. {Coates}, G.~R. {Lewis},
  D.~T. {Young}, F.~J. {Crary}, R.~A. {Baragiola}, R.~E. {Johnson}, Y.~{Dong},
  R.~J. {Wilson}, G.~H. {Jones}, J.-E. {Wahlund}, D.~G. {Mitchell}, and
  M.~{Hor{\'a}nyi}.
\newblock {Charged nanograins in the Enceladus plume}.
\newblock \emph{Journal of Geophysical Research (Space Physics)}, 117:\penalty0
  A05209, May 2012.
\newblock \doi{10.1029/2011JA017218}.

\bibitem[{Hlavenka} et~al.(2009){Hlavenka}, {Otto}, {Trippel}, {Mikosch},
  {Weidem{\"u}ller}, and {Wester}]{Hlavanka09}
P.~{Hlavenka}, R.~{Otto}, S.~{Trippel}, J.~{Mikosch}, M.~{Weidem{\"u}ller}, and
  R.~{Wester}.
\newblock {Absolute photodetachment cross section measurements of the O$^{-}$
  and OH$^{-}$ anion}.
\newblock \emph{Journal of Computational Physics}, 130\penalty0 (6):\penalty0
  061105--061105, Feb. 2009.
\newblock \doi{10.1063/1.3080809}.

\bibitem[{Hockney} and {Eastwood}(1981)]{Hockney81}
R.~W. {Hockney} and J.~W. {Eastwood}.
\newblock \emph{{Computer Simulation Using Particles}}.
\newblock CRC Press, 1981.

\bibitem[{Hoogeveen}(1994)]{Hoogeveen94}
G.~W. {Hoogeveen}.
\newblock \emph{{The Triton-Neptune Plasma Interaction.}}
\newblock PhD thesis, Rice University, Jan. 1994.

\bibitem[Huebner et~al.(1992)Huebner, Keady, and Lyon]{Huebner92}
W.~F. Huebner, J.~J. Keady, and S.~Lyon.
\newblock Solar photo rates for planetary atmospheres and atmospheric
  pollutants.
\newblock \emph{Astrophysics and Space Science}, 195\penalty0 (1):\penalty0
  1--294, 1992.

\bibitem[{Huybrighs} et~al.(2020){Huybrighs}, {Roussos}, {Bl{\"o}cker},
  {Krupp}, {Futaana}, {Barabash}, {Hadid}, {Holmberg}, {Lomax}, and
  {Witasse}]{Huybrighs}
H.~L.~F. {Huybrighs}, E.~{Roussos}, A.~{Bl{\"o}cker}, N.~{Krupp}, Y.~{Futaana},
  S.~{Barabash}, L.~Z. {Hadid}, M.~K.~G. {Holmberg}, O.~{Lomax}, and
  O.~{Witasse}.
\newblock {An Active Plume Eruption on Europa During Galileo Flyby E26 as
  Indicated by Energetic Proton Depletions}.
\newblock \emph{Geophysical Research Letters}, 47\penalty0 (10):\penalty0
  e87806, May 2020.
\newblock \doi{10.1029/2020GL087806}.

\bibitem[{Jia} et~al.(2018){Jia}, {Kivelson}, {Khurana}, and {Kurth}]{Jia18}
X.~{Jia}, M.~G. {Kivelson}, K.~K. {Khurana}, and W.~S. {Kurth}.
\newblock {Evidence of a plume on Europa from Galileo magnetic and plasma wave
  signatures}.
\newblock \emph{Nature Astronomy}, 2:\penalty0 459--464, May 2018.
\newblock \doi{10.1038/s41550-018-0450-z}.

\bibitem[Johnson et~al.(1958)Johnson, Meadows, and Holmes]{Johnson58}
C.~Y. Johnson, E.~B. Meadows, and J.~C. Holmes.
\newblock Ion composition of the arctic ionosphere.
\newblock \emph{Journal of Geophysical Research}, 63\penalty0 (2):\penalty0
  443--444, 1958.
\newblock ISSN 2156-2202.
\newblock \doi{10.1029/JZ063i002p00443}.
\newblock URL \url{http://dx.doi.org/10.1029/JZ063i002p00443}.

\bibitem[{Johnson}(1993)]{Johnson93}
R.~E. {Johnson}.
\newblock {Sputtering of an atmosphere}.
\newblock \emph{Trends in Geophys. Res.}, 2:\penalty0 501--513, 1993.
\newblock \doi{10.1029/98GL02565}.

\bibitem[{Jones} et~al.(2009){Jones}, {Arridge}, {Coates}, {Lewis}, {Kanani},
  {Wellbrock}, {Young}, {Crary}, {Tokar}, {Wilson}, {Hill}, {Johnson},
  {Mitchell}, {Schmidt}, {Kempf}, {Beckmann}, {Russell}, {Jia}, {Dougherty},
  {Waite}, and {Magee}]{Jones09}
G.~H. {Jones}, C.~S. {Arridge}, A.~J. {Coates}, G.~R. {Lewis}, S.~{Kanani},
  A.~{Wellbrock}, D.~T. {Young}, F.~J. {Crary}, R.~L. {Tokar}, R.~J. {Wilson},
  T.~W. {Hill}, R.~E. {Johnson}, D.~G. {Mitchell}, J.~{Schmidt}, S.~{Kempf},
  U.~{Beckmann}, C.~T. {Russell}, Y.~D. {Jia}, M.~K. {Dougherty}, J.~H.
  {Waite}, and B.~A. {Magee}.
\newblock {Fine jet structure of electrically charged grains in Enceladus'
  plume}.
\newblock \emph{Geophysical Research Letters}, 36:\penalty0 L16204, Aug. 2009.
\newblock \doi{10.1029/2009GL038284}.

\bibitem[{Khurana} et~al.(1998){Khurana}, {Kivelson}, {Stevenson}, {Schubert},
  {Russell}, {Walker}, and {Polanskey}]{Khurana98}
K.~K. {Khurana}, M.~G. {Kivelson}, D.~J. {Stevenson}, G.~{Schubert}, C.~T.
  {Russell}, R.~J. {Walker}, and C.~{Polanskey}.
\newblock {Induced magnetic fields as evidence for subsurface oceans in Europa
  and Callisto}.
\newblock \emph{Nature}, 395:\penalty0 777--780, Oct. 1998.
\newblock \doi{10.1038/27394}.

\bibitem[{Kivelson} et~al.(2009){Kivelson}, {Khurana}, and
  {Volwerk}]{Kivelson09}
M.~G. {Kivelson}, K.~K. {Khurana}, and M.~{Volwerk}.
\newblock \emph{{Europa's Interaction with the Jovian Magnetosphere}}, page
  545.
\newblock University of Arizona Press, 2009.

\bibitem[Kumar et~al.(2013)Kumar, Hauser, Jindra, Best, Rou{\v{c}}ka, Geppert,
  Millar, and Wester]{Kumar13}
S.~S. Kumar, D.~Hauser, R.~Jindra, T.~Best, {\v{S}}.~Rou{\v{c}}ka, W.~D.
  Geppert, T.~J. Millar, and R.~Wester.
\newblock {PHOTODETACHMENT} {AS} a {DESTRUCTION} {MECHANISM} {FOR}
  {CN}{\textendash}{AND} c3n{\textendash}{ANIONS} {IN} {CIRCUMSTELLAR}
  {ENVELOPES}.
\newblock \emph{The Astrophysical Journal}, 776\penalty0 (1):\penalty0 25, sep
  2013.
\newblock \doi{10.1088/0004-637x/776/1/25}.

\bibitem[{Ledvina} and {Brecht}(2012)]{Ledvina12}
S.~A. {Ledvina} and S.~H. {Brecht}.
\newblock {Consequences of negative ions for Titan's plasma interaction}.
\newblock \emph{Geophysical Research Letters}, 39:\penalty0 L20103, Oct. 2012.
\newblock \doi{10.1029/2012GL053835}.

\bibitem[{Lee} and {Smith}(1979)]{Lee79}
L.~C. {Lee} and G.~P. {Smith}.
\newblock {Photodissociation and photodetachment of molecular negative ions.
  VI. Ions in O$_{2}$/CH$_{4}$/H$_{2}$O mixtures from 3500 to 8600 {\r{A}}}.
\newblock \emph{Journal of Computational Physics}, 70\penalty0 (4):\penalty0
  1727--1735, Feb. 1979.
\newblock \doi{10.1063/1.437690}.

\bibitem[{Lepri} et~al.(2017){Lepri}, {Raines}, {Gilbert}, {Cutler}, {Panning},
  and {Zurbuchen}]{Lepri17}
S.~T. {Lepri}, J.~M. {Raines}, J.~A. {Gilbert}, J.~{Cutler}, M.~{Panning}, and
  T.~H. {Zurbuchen}.
\newblock {Detecting negative ions on board small satellites}.
\newblock \emph{Journal of Geophysical Research (Space Physics)}, 122\penalty0
  (4):\penalty0 3961--3971, Apr. 2017.
\newblock \doi{10.1002/2016JA023327}.

\bibitem[{Lykke} et~al.(1991){Lykke}, {Murray}, and {Lineberger}]{Lykke91}
K.~R. {Lykke}, K.~K. {Murray}, and W.~C. {Lineberger}.
\newblock {Threshold photodetachment of H$^{-}$}.
\newblock \emph{Physical Review A: General Physics}, 43\penalty0 (11):\penalty0
  6104--6107, June 1991.
\newblock \doi{10.1103/PhysRevA.43.6104}.

\bibitem[McCarthy et~al.(2006)McCarthy, Gottlieb, Gupta, and
  Thaddeus]{McCarthy06}
M.~C. McCarthy, C.~A. Gottlieb, H.~Gupta, and P.~Thaddeus.
\newblock Laboratory and astronomical identification of the negative molecular
  ion c6h–.
\newblock \emph{The Astrophysical Journal Letters}, 652\penalty0 (2):\penalty0
  L141, 2006.
\newblock URL \url{http://stacks.iop.org/1538-4357/652/i=2/a=L141}.

\bibitem[{Meeks} et~al.(2016){Meeks}, {Simon}, and {Kabanovic}]{Meeks16}
Z.~{Meeks}, S.~{Simon}, and S.~{Kabanovic}.
\newblock {A comprehensive analysis of ion cyclotron waves in the equatorial
  magnetosphere of Saturn}.
\newblock \emph{Planetary and Space Science}, 129:\penalty0 47--60, Sept. 2016.
\newblock \doi{10.1016/j.pss.2016.06.003}.

\bibitem[{Mihailescu} et~al.(2020){Mihailescu}, {Desai}, {Shebanits},
  {Haythornthwaite}, {Wellbrock}, {Coates}, {Eastwood}, and
  {Waite}]{Mihailescu20}
T.~{Mihailescu}, R.~T. {Desai}, O.~{Shebanits}, R.~{Haythornthwaite},
  A.~{Wellbrock}, A.~J. {Coates}, J.~P. {Eastwood}, and J.~H. {Waite}.
\newblock {Spatial Variations of Low-mass Negative Ions in Titan's Upper
  Atmosphere}.
\newblock \emph{The Planetary Science Journal}, 1\penalty0 (2):\penalty0 50,
  Sept. 2020.
\newblock \doi{10.3847/PSJ/abb1ba}.

\bibitem[{Mikosch} et~al.(2010){Mikosch}, {Weidem{\"u}ller}, and
  {Wester}]{Mikosh10}
J.~{Mikosch}, M.~{Weidem{\"u}ller}, and R.~{Wester}.
\newblock {On the dynamics of chemical reactions of negative ions}.
\newblock \emph{arXiv e-prints}, art. arXiv:1006.3851, June 2010.

\bibitem[{Millar} et~al.(2007){Millar}, {Walsh}, {Cordiner}, {N{\'\i}
  Chuim{\'\i}n}, and {Herbst}]{Millar07}
T.~J. {Millar}, C.~{Walsh}, M.~A. {Cordiner}, R.~{N{\'\i} Chuim{\'\i}n}, and
  E.~{Herbst}.
\newblock {Hydrocarbon Anions in Interstellar Clouds and Circumstellar
  Envelopes}.
\newblock \emph{Astrophysical Journall}, 662\penalty0 (2):\penalty0 L87--L90,
  June 2007.
\newblock \doi{10.1086/519376}.

\bibitem[Millar et~al.(2017)Millar, Walsh, and Field]{Millar17}
T.~J. Millar, C.~Walsh, and T.~A. Field.
\newblock Negative ions in space.
\newblock \emph{Chemical Reviews}, 117\penalty0 (3):\penalty0 1765--1795, 2017.
\newblock \doi{10.1021/acs.chemrev.6b00480}.
\newblock URL \url{http://dx.doi.org/10.1021/acs.chemrev.6b00480}.
\newblock PMID: 28112897.

\bibitem[{Morooka} et~al.(2011){Morooka}, {Wahlund}, {Eriksson}, {Farrell},
  {Gurnett}, {Kurth}, {Persoon}, {Shafiq}, {Andr{\'e}}, and
  {Holmberg}]{Morooka11}
M.~W. {Morooka}, J.-E. {Wahlund}, A.~I. {Eriksson}, W.~M. {Farrell}, D.~A.
  {Gurnett}, W.~S. {Kurth}, A.~M. {Persoon}, M.~{Shafiq}, M.~{Andr{\'e}}, and
  M.~K.~G. {Holmberg}.
\newblock {Dusty plasma in the vicinity of Enceladus}.
\newblock \emph{Journal of Geophysical Research (Space Physics)}, 116\penalty0
  (A15):\penalty0 A12221, Dec. 2011.
\newblock \doi{10.1029/2011JA017038}.

\bibitem[{Morooka} et~al.(2019){Morooka}, {Wahlund}, {Hadid}, {Eriksson},
  {Edberg}, {Vigren}, {Andrews}, {Persoon}, {Kurth}, {Gurnett}, {Farrell},
  {Waite}, {Perryman}, and {Perry}]{Morooka19}
M.~W. {Morooka}, J.~E. {Wahlund}, L.~Z. {Hadid}, A.~I. {Eriksson}, N.~J.~T.
  {Edberg}, E.~{Vigren}, D.~J. {Andrews}, A.~M. {Persoon}, W.~S. {Kurth}, D.~A.
  {Gurnett}, W.~M. {Farrell}, J.~H. {Waite}, R.~S. {Perryman}, and M.~{Perry}.
\newblock {Saturn's Dusty Ionosphere}.
\newblock \emph{Journal of Geophysical Research (Space Physics)}, 124\penalty0
  (3):\penalty0 1679--1697, Mar. 2019.
\newblock \doi{10.1029/2018JA026154}.

\bibitem[{Mukundan} and {Bhardwaj}(2018)]{Mukundan18}
V.~{Mukundan} and A.~{Bhardwaj}.
\newblock {A Model for Negative Ion Chemistry in Titan{\textquoteright}s
  Ionosphere}.
\newblock \emph{Astrophysical Journal}, 856\penalty0 (2):\penalty0 168, Apr.
  2018.
\newblock \doi{10.3847/1538-4357/aab1f5}.

\bibitem[{Mulliken}(1934)]{Mulliken34}
R.~S. {Mulliken}.
\newblock {A New Electroaffinity Scale; Together with Data on Valence States
  and on Valence Ionization Potentials and Electron Affinities}.
\newblock \emph{Journal of Computational Physics}, 2\penalty0 (11):\penalty0
  782--793, Nov. 1934.
\newblock \doi{10.1063/1.1749394}.

\bibitem[{Nordheim} et~al.(2020){Nordheim}, {Wellbrock}, {Jones}, {Desai},
  {Coates}, {Teolis}, and {Jasinski}]{Nordheim20}
T.~A. {Nordheim}, A.~{Wellbrock}, G.~H. {Jones}, R.~T. {Desai}, A.~J. {Coates},
  B.~D. {Teolis}, and J.~M. {Jasinski}.
\newblock {Detection of Negative Pickup Ions at Saturn's Moon Dione}.
\newblock \emph{Geophysical Research Letters}, 47\penalty0 (7):\penalty0
  e87543, Apr. 2020.
\newblock \doi{10.1029/2020GL087543}.

\bibitem[Pavlov(2014)]{Pavlov13}
A.~V. Pavlov.
\newblock Photochemistry of ions at d-region altitudes of the ionosphere: A
  review.
\newblock \emph{Surveys in Geophysics}, 35\penalty0 (2):\penalty0 259--334, Mar
  2014.
\newblock ISSN 1573-0956.
\newblock \doi{10.1007/s10712-013-9253-z}.
\newblock URL \url{https://doi.org/10.1007/s10712-013-9253-z}.

\bibitem[{Peko} and {Stephen}(2000)]{Peko00}
B.~L. {Peko} and T.~M. {Stephen}.
\newblock {Absolute detection efficiencies of low energy H, H $^{-}$, H $^{+}$,
  H $_{2}$$^{+}$ and H $_{3}$$^{+}$ incident on a multichannel plate detector}.
\newblock \emph{Nuclear Instruments and Methods in Physics Research B},
  171:\penalty0 597--604, Dec. 2000.
\newblock \doi{10.1016/S0168-583X(00)00306-2}.

\bibitem[{Phillips} and {Pappalardo}(2014)]{Phillips14}
C.~B. {Phillips} and R.~T. {Pappalardo}.
\newblock {Europa Clipper Mission Concept: Exploring Jupiter's Ocean Moon}.
\newblock \emph{EOS Transactions}, 95\penalty0 (20):\penalty0 165--167, May
  2014.
\newblock \doi{10.1002/2014EO200002}.

\bibitem[{Regoli} et~al.(2016){Regoli}, {Coates}, {Thomsen}, {Jones},
  {Roussos}, {Waite}, {Krupp}, and {Cox}]{Regoli16}
L.~H. {Regoli}, A.~J. {Coates}, M.~F. {Thomsen}, G.~H. {Jones}, E.~{Roussos},
  J.~H. {Waite}, N.~{Krupp}, and G.~{Cox}.
\newblock {Survey of pickup ion signatures in the vicinity of Titan using
  CAPS/IMS}.
\newblock \emph{Journal of Geophysical Research (Space Physics)}, 121:\penalty0
  8317--8328, Sept. 2016.
\newblock \doi{10.1002/2016JA022617}.

\bibitem[{Robinson} and {Geltman}(1967)]{Robinson67}
E.~J. {Robinson} and S.~{Geltman}.
\newblock {Single- and Double-Quantum Photodetachment of Negative Ions}.
\newblock \emph{Physical Review}, 153\penalty0 (1):\penalty0 4--8, Jan. 1967.
\newblock \doi{10.1103/PhysRev.153.4}.

\bibitem[{Roth} et~al.(2014){Roth}, {Saur}, {Retherford}, {Strobel}, {Feldman},
  {McGrath}, and {Nimmo}]{Roth14}
L.~{Roth}, J.~{Saur}, K.~D. {Retherford}, D.~F. {Strobel}, P.~D. {Feldman},
  M.~A. {McGrath}, and F.~{Nimmo}.
\newblock {Transient Water Vapor at Europa's South Pole}.
\newblock \emph{Science}, 343\penalty0 (6167):\penalty0 171--174, Jan. 2014.
\newblock \doi{10.1126/science.1247051}.

\bibitem[{Scipioni} et~al.(2014){Scipioni}, {Tosi}, {Stephan}, {Filacchione},
  {Ciarniello}, {Capaccioni}, and {Cerroni}]{Scipioni14}
F.~{Scipioni}, F.~{Tosi}, K.~{Stephan}, G.~{Filacchione}, M.~{Ciarniello},
  F.~{Capaccioni}, and P.~{Cerroni}.
\newblock {Spectroscopic classification of icy satellites of Saturn II:
  Identification of terrain units on Rhea}.
\newblock \emph{Icarus}, 234:\penalty0 1--16, May 2014.
\newblock \doi{10.1016/j.icarus.2014.02.010}.

\bibitem[{Seman} and {Branscomb}(1962)]{Seman62}
M.~L. {Seman} and L.~M. {Branscomb}.
\newblock {Structure and Photodetachment Spectrum of the Atomic Carbon Negative
  Ion}.
\newblock \emph{Physical Review}, 125\penalty0 (5):\penalty0 1602--1608, Mar.
  1962.
\newblock \doi{10.1103/PhysRev.125.1602}.

\bibitem[{Shebanits} et~al.(2013){Shebanits}, {Wahlund}, {Mandt}, {{\AA}gren},
  {Edberg}, and {Waite}]{Shebanits13}
O.~{Shebanits}, J.-E. {Wahlund}, K.~{Mandt}, K.~{{\AA}gren}, N.~J.~T. {Edberg},
  and J.~H. {Waite}.
\newblock {Negative ion densities in the ionosphere of Titan-Cassini RPWS/LP
  results}.
\newblock \emph{Planetary and Space Science}, 84:\penalty0 153--162, Aug. 2013.
\newblock \doi{10.1016/j.pss.2013.05.021}.

\bibitem[{Shebanits} et~al.(2016){Shebanits}, {Wahlund}, {Edberg}, {Crary},
  {Wellbrock}, {Andrews}, {Vigren}, {Desai}, {Coates}, {Mandt}, and
  {Waite}]{Shebanits16}
O.~{Shebanits}, J.-E. {Wahlund}, N.~J.~T. {Edberg}, F.~J. {Crary},
  A.~{Wellbrock}, D.~J. {Andrews}, E.~{Vigren}, R.~T. {Desai}, A.~J. {Coates},
  K.~E. {Mandt}, and J.~H. {Waite}.
\newblock {Ion and aerosol precursor densities in Titan's ionosphere: A
  multi-instrument case study}.
\newblock \emph{Journal of Geophysical Research (Space Physics)}, 121\penalty0
  (A10):\penalty0 10, Oct. 2016.
\newblock \doi{10.1002/2016JA022980}.

\bibitem[{Shebanits} et~al.(2020){Shebanits}, {Hadid}, {Cao}, {Morooka},
  {Hunt}, {Dougherty}, {Wahlund}, {Waite}, and
  {M{\"u}ller-Wodarg}]{Shebanits20}
O.~{Shebanits}, L.~Z. {Hadid}, H.~{Cao}, M.~W. {Morooka}, G.~J. {Hunt}, M.~K.
  {Dougherty}, J.~E. {Wahlund}, J.~H. {Waite}, and I.~{M{\"u}ller-Wodarg}.
\newblock {Saturn's near-equatorial ionospheric conductivities from in situ
  measurements}.
\newblock \emph{Scientific Reports}, 10:\penalty0 7932, May 2020.
\newblock \doi{10.1038/s41598-020-64787-7}.

\bibitem[{Sparks} et~al.(2016){Sparks}, {Hand}, {McGrath}, {Bergeron},
  {Cracraft}, and {Deustua}]{Sparks16}
W.~B. {Sparks}, K.~P. {Hand}, M.~A. {McGrath}, E.~{Bergeron}, M.~{Cracraft},
  and S.~E. {Deustua}.
\newblock {Probing for Evidence of Plumes on Europa with HST/STIS}.
\newblock \emph{Astrophysical Journal}, 829\penalty0 (2):\penalty0 121, Oct.
  2016.
\newblock \doi{10.3847/0004-637X/829/2/121}.

\bibitem[Stephen and Peko(2000)]{Stephen00}
T.~M. Stephen and B.~L. Peko.
\newblock Absolute calibration of a multichannel plate detector for low energy
  o, o$^-$, and o$^+$.
\newblock \emph{Review of Scientific Instruments}, 71\penalty0 (3):\penalty0
  1355--1359, 2000.
\newblock \doi{10.1063/1.1150462}.
\newblock URL \url{http://dx.doi.org/10.1063/1.1150462}.

\bibitem[Taylor et~al.(2018)Taylor, Coates, Jones, Wellbrock, Fazakerley,
  Desai, Caro-Carretero, Michiko, Schippers, and Waite]{Taylor17}
S.~A. Taylor, A.~J. Coates, G.~H. Jones, A.~Wellbrock, A.~N. Fazakerley, R.~T.
  Desai, R.~Caro-Carretero, M.~W. Michiko, P.~Schippers, and J.~H. Waite.
\newblock Modeling, analysis, and interpretation of photoelectron energy
  spectra at enceladus observed by cassini.
\newblock \emph{Journal of Geophysical Research: Space Physics}, 123\penalty0
  (1):\penalty0 287--296, 2018.
\newblock ISSN 2169-9402.
\newblock \doi{10.1002/2017JA024536}.
\newblock URL \url{http://dx.doi.org/10.1002/2017JA024536}.
\newblock 2017JA024536.

\bibitem[{Teolis} et~al.(2010){Teolis}, {Jones}, {Miles}, {Tokar}, {Magee},
  {Waite}, {Roussos}, {Young}, {Crary}, {Coates}, {Johnson}, {Tseng}, and
  {Baragiola}]{Teolis10}
B.~D. {Teolis}, G.~H. {Jones}, P.~F. {Miles}, R.~L. {Tokar}, B.~A. {Magee},
  J.~H. {Waite}, E.~{Roussos}, D.~T. {Young}, F.~J. {Crary}, A.~J. {Coates},
  R.~E. {Johnson}, W.-L. {Tseng}, and R.~A. {Baragiola}.
\newblock {Cassini Finds an Oxygen-Carbon Dioxide Atmosphere at Saturn’s Icy
  Moon Rhea}.
\newblock \emph{Science}, 330:\penalty0 1813, Dec. 2010.
\newblock \doi{10.1126/science.1198366}.

\bibitem[{Tokar} et~al.(2012){Tokar}, {Johnson}, {Thomsen}, {Sittler},
  {Coates}, {Wilson}, {Crary}, {Young}, and {Jones}]{Tokar12}
R.~L. {Tokar}, R.~E. {Johnson}, M.~F. {Thomsen}, E.~C. {Sittler}, A.~J.
  {Coates}, R.~J. {Wilson}, F.~J. {Crary}, D.~T. {Young}, and G.~H. {Jones}.
\newblock {Detection of exospheric O$_{2}$$^{+}$ at Saturn's moon Dione}.
\newblock \emph{Geophysical Research Letters}, 39:\penalty0 L03105, Feb. 2012.
\newblock \doi{10.1029/2011GL050452}.

\bibitem[{Van Allen} et~al.(1975){Van Allen}, {Randall}, {Baker}, {Goertz},
  {Sentman}, {Thomsen}, and {Flindt}]{VanAllen75}
J.~A. {Van Allen}, B.~A. {Randall}, D.~N. {Baker}, C.~K. {Goertz}, D.~D.
  {Sentman}, M.~F. {Thomsen}, and H.~R. {Flindt}.
\newblock {Pioneer 11 Observations of Energetic Particles in the Jovian
  Magnetosphere}.
\newblock \emph{Science}, 188\penalty0 (4187):\penalty0 459--462, May 1975.
\newblock \doi{10.1126/science.188.4187.459}.

\bibitem[{Volwerk} et~al.(2001){Volwerk}, {Kivelson}, and {Khurana}]{Volwerk01}
M.~{Volwerk}, M.~G. {Kivelson}, and K.~K. {Khurana}.
\newblock {Wave activity in Europa's wake: Implications for ion pickup}.
\newblock \emph{Journal of Geophysical Research}, 106:\penalty0 26033--26048,
  Nov. 2001.
\newblock \doi{10.1029/2000JA000347}.

\bibitem[{Vuitton} et~al.(2009){Vuitton}, {Lavvas}, {Yelle}, {Galand},
  {Wellbrock}, {Lewis}, {Coates}, and {Wahlund}]{Vuitton09}
V.~{Vuitton}, P.~{Lavvas}, R.~V. {Yelle}, M.~{Galand}, A.~{Wellbrock}, G.~R.
  {Lewis}, A.~J. {Coates}, and J.-E. {Wahlund}.
\newblock {Negative ion chemistry in Titan's upper atmosphere}.
\newblock \emph{Planetary and Space Science}, 57:\penalty0 1558--1572, Nov.
  2009.
\newblock \doi{10.1016/j.pss.2009.04.004}.

\bibitem[{Wahlund} et~al.(2009){Wahlund}, {Galand}, {M{\"u}ller-Wodarg}, {Cui},
  {Yelle}, {Crary}, {Mandt}, {Magee}, {Waite}, {Young}, {Coates}, {Garnier},
  {{\AA}gren}, {Andr{\'e}}, {Eriksson}, {Cravens}, {Vuitton}, {Gurnett}, and
  {Kurth}]{Wahlund09}
J.-E. {Wahlund}, M.~{Galand}, I.~{M{\"u}ller-Wodarg}, J.~{Cui}, R.~V. {Yelle},
  F.~J. {Crary}, K.~{Mandt}, B.~{Magee}, J.~H. {Waite}, D.~T. {Young}, A.~J.
  {Coates}, P.~{Garnier}, K.~{{\AA}gren}, M.~{Andr{\'e}}, A.~I. {Eriksson},
  T.~E. {Cravens}, V.~{Vuitton}, D.~A. {Gurnett}, and W.~S. {Kurth}.
\newblock {On the amount of heavy molecular ions in Titan's ionosphere}.
\newblock \emph{Planetary and Space Science}, 57:\penalty0 1857--1865, Dec.
  2009.
\newblock \doi{10.1016/j.pss.2009.07.014}.

\bibitem[{Wekhof}(1981)]{Wekhof81}
A.~{Wekhof}.
\newblock {Negative ions in comets}.
\newblock \emph{Moon and Planets}, 24:\penalty0 157--173, Apr. 1981.
\newblock \doi{10.1007/BF00910606}.

\bibitem[{Wellbrock} et~al.(2013){Wellbrock}, {Coates}, {Jones}, {Lewis}, and
  {Waite}]{Wellbrock13}
A.~{Wellbrock}, A.~J. {Coates}, G.~H. {Jones}, G.~R. {Lewis}, and J.~H.
  {Waite}.
\newblock {Cassini CAPS-ELS observations of negative ions in Titan's
  ionosphere: Trends of density with altitude}.
\newblock \emph{Geophysical Research Letters}, 40:\penalty0 4481--4485, Sept.
  2013.
\newblock \doi{10.1002/grl.50751}.

\bibitem[{Wellbrock} et~al.(2019){Wellbrock}, {Coates}, {Jones}, {Vuitton},
  {Lavvas}, {Desai}, and {Waite}]{Wellbrock19}
A.~{Wellbrock}, A.~J. {Coates}, G.~H. {Jones}, V.~{Vuitton}, P.~{Lavvas}, R.~T.
  {Desai}, and J.~H. {Waite}.
\newblock {Heavy negative ion growth in Titan's polar winter}.
\newblock \emph{Monthly Notices of the Royal Astronomical Society},
  490\penalty0 (2):\penalty0 2254--2261, Dec. 2019.
\newblock \doi{10.1093/mnras/stz2655}.

\bibitem[{Wildt}(1939)]{Wildt39}
R.~{Wildt}.
\newblock {Negative Ions of Hydrogen and the Opacity of Stellar Atmospheres.}
\newblock \emph{Astrophysical Journal}, 90:\penalty0 611, Nov. 1939.
\newblock \doi{10.1086/144125}.

\bibitem[{Wilson} et~al.(2010){Wilson}, {Tokar}, {Kurth}, and
  {Persoon}]{Wilson10}
R.~J. {Wilson}, R.~L. {Tokar}, W.~S. {Kurth}, and A.~M. {Persoon}.
\newblock {Properties of the thermal ion plasma near Rhea as measured by the
  Cassini plasma spectrometer}.
\newblock \emph{Journal of Geophysical Research (Space Physics)}, 115:\penalty0
  A05201, May 2010.
\newblock \doi{10.1029/2009JA014679}.

\bibitem[{Wishart}(1979)]{Wishart79}
A.~W. {Wishart}.
\newblock {The bound-free photodetachment cross section of H$^{-}$}.
\newblock \emph{Journal of Physics B Atomic Molecular Physics}, 12\penalty0
  (21):\penalty0 3511--3519, Nov. 1979.
\newblock \doi{10.1088/0022-3700/12/21/009}.

\bibitem[Zhang et~al.(2021)Zhang, Desai, Miyake, Usui, and Shebanits]{Zhang21}
Z.~Zhang, R.~T. Desai, Y.~Miyake, H.~Usui, and O.~Shebanits.
\newblock {Particle-in-cell simulations of the Cassini spacecraft’s
  interaction with Saturn’s ionosphere during the Grand Finale}.
\newblock \emph{Monthly Notices of the Royal Astronomical Society},
  504\penalty0 (1):\penalty0 964--973, 03 2021.
\newblock ISSN 0035-8711.
\newblock \doi{10.1093/mnras/stab750}.
\newblock URL \url{https://doi.org/10.1093/mnras/stab750}.

\bibitem[{Zhang} et~al.(2021, April 7){Zhang}, {Wu}, and {Desai}]{ZhangData}
Z.~{Zhang}, X.~{Wu}, and R.~{Desai}.
\newblock {CubeOvO/Empirical-Photodetachment-Rate: Updated (Version v2.0)},
  Apr. 2021, April 7.
\newblock URL \url{http://doi.org/10.5281/zenodo.4670382}.

\end{thebibliography}

\end{document}